\documentclass[reprint,floats,floatfix,amsmath,amssymb,nofootinbib,longbibliography]{revtex4-2}
\usepackage{amsfonts}
\usepackage{graphicx}
\usepackage{dcolumn}
\usepackage{bm}
\usepackage{natbib}
\usepackage{wasysym}
\usepackage{multirow}
\usepackage{aas_macros}
\usepackage[hyperindex,colorlinks]{hyperref}

\newcommand{\bs}{\boldsymbol}

\newcommand{\G}{\mathcal{G}^{(4)}}
\newcommand{\PhiG}{\Phi_{\mathcal{G}}}

\begin{document}

\title{\textbf{Exploring Scalarized Einstein-Gauss-Bonnet Theories Through the Lens of Parametrized Post-Newtonian Formalism} 
}%

\author{Mart\'in G. Richarte}
\email{martin@df.uba.ar}
\affiliation{Departamento de F\'isica - Universidade Federal do Esp\'irito Santo, 29075-910 Vit\'oria, ES, Brazil}
\affiliation{PPGCosmo, CCE - Universidade Federal do Esp\'irito Santo, 29075-910 Vit\'oria, ES, Brazil}
\affiliation{Departamento de F\'isica, Facultad de Ciencias Exactas y Naturales,
Universidad de Buenos Aires, Ciudad Universitaria 1428, Pabell\'on I, Buenos Aires, Argentina}

\author{Júnior D. Toniato}
 \email{junior.toniato@ufes.br}
\affiliation{Departamento de Química e Física - Centro de Ciências Exatas, Naturais e da Saúde, Universidade Federal do Espírito Santo - Campus Alegre, ES, 29500-000, Brazil.}%
\affiliation{Núcleo de Astrofísica e Cosmologia - Cosmo-Ufes, Universidade Federal do Espírito Santo, Vitória, ES,  29075-910, Brazil.}

\date{\today}

\begin{abstract}
We explore scalarized Einstein-Gauss-Bonnet  theories within the context of the Parameterized Post-Newtonian formalism, which serves as a robust framework for examining modifications to General Relativity that exhibit scalarization. This approach enables us to impose a variety of constraints on the parameter space, particularly focusing on the PPN parameters $\gamma$ and $\beta$. Our analysis reveals several significant bounds for each scalarized model, utilizing data from the Cassini mission, Lunar Laser Ranging, and the forthcoming BepiColombo mission, in conjunction with geodetic Very Long Baseline Interferometry measurements around the Sun. In particular, we conducted a combined analysis of the MESSENGER mission's precision measurements of Mercury's perihelion precession $\dot{\omega}$ alongside the Cassini constraints on $\gamma$, yielding intriguing limits for the Ricci-EGB model. Furthermore, we investigate the impact of the new post-Newtonian potential introduced by the Gauss-Bonnet term at fourth order, which necessitates an expanded formulation of the PPN parameters. This highlights that the parameter $\beta$ exhibits different effects when analyzing the Ricci-EGB model. Additional constraints are derived from this framework by estimating the $\gamma$ PPN parameter using data from strong-lensing galaxy systems.

\end{abstract}

\maketitle

\section{Introduction}
The scalarization mechanism was first addressed by Damour and Esposito-Farése in 1993, within the framework of scalar tensor theories \cite{Damour:1993hw}. The underlying principle of the DEF model is that by introducing a coupling function between the metric field and an evolving scalar field at the level of the matter action, the new model closely resembles General Relativity (GR) in the weak-field limit and low-velocity regime, while diverging from GR in the strong-field context. To demonstrate this concept, they illustrate that the DEF scenario for neutron stars in the post-Keplerian approximation may facilitate the detection of contributions from the scalar field degree of freedom.  Since then, theories that exhibit a scalarization mechanism have been explored extensively (see \cite{Doneva:2022ewd} for a recent review on this topic). In particular, theories incorporating a quadratic Gauss-Bonnet term have gained significant attention in the context of hairy black holes \cite{Antoniou:2017acq, Herdeiro:2018wub, Dima:2020yac, East:2021bqk, Berti:2020kgk, Zhang:2021nnn, Silva:2020omi, Herdeiro:2020wei, Cunha:2019dwb} or compact stars \cite{Silva:2017uqg, Mendes:2018qwo, Kuan:2021lol}. 

Theoretical motivations for exploring these models are several. For example, they represent a somewhat natural extension of GR by introducing a new degree of freedom \cite{Clifton:2011jh, Koyama:2015vza, Joyce:2014kja, Ishak:2018his}. Furthermore, they enrich the phenomenology by incorporating a coupling function between the scalar field and the Gauss-Bonnet term, which facilitates the onset of tachyon instability at the linear level. Another reason is that spontaneous scalarization does not require the existence of additional matter fields, as is the case with the traditional DEF model \cite{Doneva:2017bvd}. 

From a physical perspective, the novel features of these models lie in their ability to circumvent the no-hair theorem; thus, asymptotically flat black hole solutions become admissible, provided that the coupling function satisfies certain conditions \cite{Antoniou:2017acq}. A pivotal aspect of the spontaneous scalarization mechanism for black hole solutions is that the inclusion of the scalar field facilitates configurations where the Schwarzschild solution not only becomes scalarized, but also exhibits enhanced thermodynamic stability compared to the scenario where the scalar field is absent \cite{Doneva:2017bvd}. Commencing with the scalar-Einstein-Gauss-Bonnet model along with a specific coupling function and focusing solely on the perturbation of the scalar field, it was found that the additional scalar field facilitates stable black hole configurations that remain inaccessible when the scalar field is effectively switched off \cite{Doneva:2017bvd}. 

It is crucial to emphasize that the scalarization mechanism for black hole solutions or compact objects is not restricted exclusively to scalar fields. This intriguing phenomenon can also be realized with various other field types, including vector fields \cite{Ramazanoglu:2017xbl, Annulli:2019fzq, Minamitsuji:2020pak, Kase:2020yhw, Ye:2024pyy, Oliveira:2020dru} and tensor fields \cite{Ramazanoglu:2019gbz}. In fact, by extending the class of disformal couplings in the matter sector, one improves the emergence of this mechanism across a spectrum of fields, which includes the vector, spinor, and spin-2 fields as well \cite{Ramazanoglu:2019jrr}. Nevertheless, the vectorization/tensorization case is not as rich as the scalar case, as it may introduce ghost instabilities in addition to tachyonic ones. This, in turn, leads to a poorly posed problem for the dynamical evolution of the vector/tensor field\cite{Silva:2021jya}. With that in mind, another avenue for developing scalarization involves coupling the scalar field to a Born-Infeld Lagrangian \cite{Wang:2020ohb}, exploring Einstein-Maxwell scalar theories with a quasitopological term \cite{Myung:2020ctt}, or considering an interaction term between the axion field and a Chern-Simons density \cite{Chatzifotis:2022mob}.

In this paper, we investigate a generalized scalar-tensor model that encompasses several of the scenarios mentioned above. Our aim is to compute the PPN parameters for these classes of theories and to examine various observational constraints, which will include not only the local solar system but also other regimes and scales. The structure of the paper is as follows. We begin by introducing the model and its various branches, deriving the field equations for both the metric and the scalar fields. Subsequently, we go into the details of how the post-Newtonian expansion operates within this generalized framework. A comprehensive analysis of the observational constraints is performed for the different PPN parameters and for the parameters associated with each model.
To be more precise, we start by outlining several constraints on the $\beta_1$ and $\gamma$ parameters derived from data obtained through the Cassini mission, Lunar Laser Ranging, geodetic VLBI measurements around the Sun, the perihelion shift of Mercury, planetary ephemerides, and the forthcoming BepiColombo mission, and the bounds coming from the MESSENGER mission on the precision Mercury's perihelion.  Additionally, we highlight bounds informed by strong galaxy-lensed systems and conduct our own Bayesian analysis for several strong-lensing systems. 
Finally, the conclusions are stated.   We use geometrized units, where $G =1$, and the metric signature is mostly $+2$.

\section{Scalarized Einstein-Gauss-Bonnet theories}\label{sec:scalarEGB}
Our starting point is to consider a generalized scalar-tensor model. The action encompasses a scalar field, denoted as $\Phi$, along with a metric field $g_{\mu\nu}$:  
\begin{equation}\label{ac1}
    S=\int{ \frac{d\mu_{g}}{2\kappa}\left[M(\Phi)R-Q(\Phi)\nabla_a\Phi\nabla^a\Phi +\frac{1}{8}\,\varepsilon f(\Phi){\cal G}\right]}.~~
\end{equation}
Here, the volume element is given by  $d\mu_{g}=\sqrt{-g} d^4x $, $R$ denotes the Ricci scalar, while ${\mathcal{G}} = R^2 + R_{abcd} R^{abcd} - 4 R_{ab} R^{ab}$ represents the quadratic Gauss-Bonnet term. The functions $M(\Phi)$, $Q(\Phi)$, and $f(\Phi)$ are arbitrary (coupling)functions of the scalar field $\Phi$. The symbol $\varepsilon$ is just a constant, and we define $\kappa = 8\pi G/c^{4}$. Alongside the action given in (\ref{ac1}), the model may also incorporate standard matter content represented as $S_{\text{matt}}[g_{\mu\nu}, \Psi]$, where $\Psi$ signifies additional fields/fluids. In the context of the DEF model, it could furthermore include a non-trivial coupling characterized by a function $A(\Phi)$, expressed as $S_{\text{matt}}[A(\Phi)g_{\mu\nu}, \Psi]$ \cite{Damour:1993hw}. By varying the action with respect to the metric and performing the necessary simplifications, we arrive at the following field equation for the metric:
\begin{align}\label{fe}
&M(\Phi) G_{\alpha \beta }- Q(\Phi) \left({\nabla}_{\alpha }\Phi {\nabla}_{\beta }\Phi + Xg_{\alpha \beta } \right) \ + \nonumber \\ \nonumber \\
&\varepsilon f'(\Phi) \left[-\frac{R}{2}{\nabla}_{\beta }{\nabla}_{\alpha }\Phi + R_{(\alpha \gamma } {\nabla}^{\gamma}{\nabla}_{\beta)}\Phi - G_{\alpha \beta } \Box\Phi \right]- \nonumber\\ \nonumber \\
&\varepsilon f'(\Phi) \left[ g_{\alpha \beta } R_{\gamma \delta } {\nabla}^{\gamma }{\nabla}^{\delta}\Phi + {R}_{\alpha \gamma \beta \delta }{\nabla}^{\gamma }{\nabla}^{\delta}\Phi \right] \ + \nonumber\\ \nonumber \\
&\varepsilon f''(\Phi) \bigg[-\frac{R}{2}{\nabla}_{\alpha }\Phi{\nabla}_{\beta }\Phi + R_{(\alpha \gamma } {\nabla}^{\gamma}\Phi{\nabla}_{\beta)}\Phi\bigg]+ \nonumber\\ \nonumber \\
&\varepsilon f''(\Phi) \bigg[
2XG_{\alpha \beta } - g_{\alpha\beta}R_{\gamma \delta }{\nabla}^{\gamma}\Phi{\nabla}^{\delta}\Phi +{R}_{\alpha \gamma \beta \delta}{\nabla}^{\gamma }\Phi{\nabla}^{\delta}\Phi \bigg] \ - \nonumber \\ \nonumber \\
&M'(\Phi) \left({\nabla}_{\beta }{\nabla}_{\alpha }\Phi -  g_{\alpha \beta } \Box\Phi \right) - M''(\Phi) {\nabla}_{\alpha }\Phi {\nabla}_{\beta }\Phi + \nonumber \\
\nonumber \\
&2M''(\Phi)X g_{\alpha \beta } =\kappa T_{\alpha \beta },
\end{align}
where $G_{\alpha\beta}$ is the Einstein tensor and $X = -\tfrac{1}{2}\nabla^{c}\Phi\nabla_{c}\Phi$. In this context, the prime symbol denotes a derivative with respect to the scalar field, whereas the indices enclosed in parentheses indicate symmetrization. It should be noted that the aforementioned field equations (\ref{fe}) can formally be expressed in terms of the double dual Riemann tensor. However, this compact form is not suitable for future use. Taking the trace of Eq. (\ref{fe}), the latter equation can be expressed as follows:
\begin{align}
  &M(\Phi) R + 2Q(\Phi)X  - \varepsilon f'(\Phi)\Big(\frac{R}{2} \Box\Phi - R_{ab} \nabla^{b}\nabla^{a}\Phi\Big) +\nonumber \\
  &\varepsilon f''(\Phi)\Big(RX + R_{ab} \nabla^{a}\Phi\nabla^{b}\Phi\Big) 
  -3 M'(\Phi)\Box\Phi + \nonumber \\ 
  &6 M''(\Phi) X = - \kappa T,
\end{align}
where $T$ denotes the trace of TEM.
The generalized scalar field equation is given by 
\begin{align}
 Q(\Phi)\Box\Phi - Q'(\Phi)X +  M'(\Phi)\frac{R}{2} + \frac{\varepsilon}{8} f'(\Phi){\cal G}=0.
\end{align}
Depending on the choice of the coupling functions mentioned above, the field equations exhibit different limits. For the sake of completeness, we shall outline some of these theories:
\begin{enumerate}
    \item For $M(\Phi)=Q(\phi)=1$ and $T=0$, we recover the scalarized EGB model \cite{Doneva:2017bvd}. There are several possibilities for $f$ coupling: $f_{1}(\Phi)=\alpha\Phi^{2}$, $f_{2}(\Phi)=\alpha\Phi^{2} + \mu \Phi^4$, $f_{3}(\Phi)=\alpha^{2}[1-e^{\beta\Phi^{2}}]$ (cf. \cite{Doneva:2022ewd}).
    
    \item For $M(\Phi)=1-\frac{\beta}{2}\Phi^{2}$, $Q(\phi)=1$, and $f(\Phi)=\alpha \Phi/2$, we obtain the Ricci-EGB scalar model \cite{Antoniou:2020nax}. 
    
    \item Taking $M(\Phi)=\Phi$, $Q(\Phi)=\omega/\Phi$, and $f(\Phi)=0$, the Brans-Dicke model is obtained.

    \item An equivalent DEF model is achieved by selecting $M(\Phi)=1-\frac{\beta}{4}\Phi^2$, $Q=1$, and $f=0$ \cite{Andreou:2019ikc}. In the original DEF proposal, $M=Q=1$ and $f=0$. However, a coupling is introduced in the action of matter through an effective metric given by $\hat{g}_{\mu\nu}=g_{\mu\nu} A^{2}(\Phi)$, where $A^{2}(\Phi)=e^{\beta \Phi^{2}/2}$.  This yields the following master equation, $\Box \Phi=-k \beta \Phi A^{4}(\Phi) T$, where $T$ is the trace of the TEM. The latter equation indicates that, in the DEF model, it is not possible to trigger the tachyon instability in the absence of a source.

    \item The case with $M(\Phi)=Q=1$ and $f(\Phi)=\alpha \Phi$ leads to $\mathcal{L}=R + X + \alpha \Phi {\cal G}$, which corresponds to a shift-symmetric Hordensky/Galileon model with $G_{2}=G_{3}=G_{4}=0$ but $G_{5}=-4\ln |X|$ \cite{Babichev:2017guv}. 
\end{enumerate}

Although the extended model we are presenting shares certain similarities with the regularized EGB theory, particularly in the presence of a Gauss-Bonnet term, the coupling functions differ significantly. Moreover, the non-linear terms in the derivatives of the scalar field, such as $\Box \Phi (\nabla \Phi)^{2}$ and $(\nabla \Phi)^{4}$, are absent \cite{Toniato:2024gtx}. Furthermore, it is important to note that the latter model does not encompass the Brans-Dicke theory or the DEF model as potential limits. In fact, the scalarRicci-EGB model is entirely excluded from consideration. Thus, we find ourselves investigating a genuine class of theories that not only encompasses many well-known models, but also extends scalarized theories by permitting a combination of couplings between the kinetic scalar field in the presence of the Gauss-Bonnet term.

In the following section, we will go into the details of computing the field equations as an expansion in powers of $v/c$ within the weak-field approach \cite{Will:1993ns},\cite{PoissonWill}.

\section{Post-Newtonian formalism in a nutshell}
To obtain the post-Newtonian parameters associated with the model (\ref{fe}), we must adhere to the parameterized post-Newtonian formalism outlined in Refs. \cite{Will:1993ns,PoissonWill}. In this endeavor, we consider a slight deviation from the Minkowski's background $\eta_{\mu\nu}$ and a constant value for the scalar field $\phi_0$:
\begin{align}\label{metricexp}
g_{\mu\nu}&=\eta_{\mu\nu}+h_{\mu\nu}\,,\\
\Phi&=\phi_0+\phi.
\end{align}
In doing so, we assume that the source matter is slow-moving ($v/c\ll1$) and weak gravitational fields ($|h_{\mu\nu}|\sim \psi_{N}\ll1$).  We also disregard the evolution of the background, which renders the PPN formalism unsuitable for cosmological applications.

Concerning the additional matter in the theory under study, we assume it can be characterized as a perfect fluid, whose energy-momentum tensor is expressed as follows:
\begin{align}\label{tem}
T^{\mu\nu}=\left(\rho+\rho\Pi+p\right)u^{\mu}u^{\nu}+pg^{\mu\nu}. 
\end{align}
In Eq. (\ref{tem}), the physical quantities include the mass density denoted by $\rho$, the internal energy of the fluid per unit mass $\Pi$, the pressure of the fluid $p$, its four-velocity $u^{\mu}=u^{0}(1, v^{i})$, and its three-velocity is $v^{i}$.  The energy-momentum tensor (\ref{tem}) is adequate to derive the post-Newtonian expansion of the gravitational field surrounding a fluid body, such as the Sun, or within a compact binary system \cite{PoissonWill}. When expanding the energy-momentum tensor in powers of $v/c$, we find that the equation of state and the internal energy are of order two, while the time derivative is of order one; specifically, $p/\rho \sim \Pi \sim (v/c)^{2} \sim \psi_{N}$ and $\partial_t \sim (v/c)$ (quasi-static hypothesis) \cite{PoissonWill}. The physical reasoning underpinning this scheme is that for local systems, the gravitational field $\psi_{N}$ varies on a time scale that is much longer than the typical orbital period \cite{Will:1971zza}. Hence, the components of the TEM are given by 
\begin{align}\label{expa3}
T_{00}&=\rho[1-h^{(2)}_{00} + \Pi + v^{2}] +\mathcal{O}(6),   \\ 
T_{0i}&=-\rho v_{i}+\mathcal{O}(5),   \\
T_{ij}&=\rho v_{i}v_{j} + p \delta_{ij}+ \mathcal{O}(6).
\end{align}

With respect to the metric expansion, the first post-Newtonian corrections to the equations of motion for massive bodies require an expansion of $h_{00}$ up to order $(v/c)^{4}$, $h_{0i}$ up to order $(v/c)^{3}$, and $h_{ij}$ up to order $(v/c)^{2}$. The generic expansion of the metric  components  and scalar field read
\begin{align}\label{expa2}
g_{00}&=-1 + h^{(2)}_{00} +  h^{(4)}_{00} + \mathcal{O}(6),   \\ 
g_{0i}&=h^{(3)}_{0i} +  h^{(5)}_{0i} + \mathcal{O}(5),\\
g_{ij}&=\delta_{ij} +  h^{(2)}_{ij} + \mathcal{O}(4),\\
\Phi&=\phi_0+ \phi_{2} + \phi_4 +\mathcal{O}(6),
\end{align}
where we assumed that the scalar field only admits an even expansion in power of $v/c$, a common assumption within the PPN framework. Moreover, any scalar function of $\Phi$ will be assumed to be Taylor expanded around the background value $\phi_0$. We are going to use the following  notation:
\begin{equation}
    F(\Phi)= F_0 + F_1\phi + F_2\phi^2 + \mathcal{O}(6).
\end{equation}

As mentioned above, the field equations are expanded in powers of $v/c$, and solved order by order to obtain the metric components \cite{PoissonWill}.  We start by solving the second-order equations,
\begin{align}
    &M_0\nabla^2h^{(2)}_{00} -M_1\nabla^2\phi_2=-\kappa\rho,\\[1ex]
    &(2M_0Q_0+3M_1^2)\nabla^2\phi_2= -M_1\kappa\rho,\\[1ex]
    M_0\Big(\nabla^2h^{(2)}_{ij} &-h^{(2)}_{00,ij}\Big) +M_0\delta^{kl}\Big(h^{(2)}_{kl,ij} -h^{(2)}_{(ik,lj)}\Big) \notag\\
&+M_1\Big(2\phi_2{}_{,ij}+\nabla^2\phi_2\delta_{ij}\Big) = -\kappa\rho\delta_{ij}.\label{eq-hij2}
\end{align}
It is easily to see that $h^{(2)}_{00}$ and $\phi_2$ are proportional to the Newtonian potential $U$, defined by
\begin{equation}\label{new1}
    \nabla^2U=-4\pi\rho,
\end{equation}
or, equivalently,
\begin{equation}\label{U}
    U=\frac{1}{4\pi}\int\dfrac{\rho{}'\,d^{3}x' }{|\bs x- \bs x'|}\,.
\end{equation}
In Eq. \eqref{eq-hij2}, we assume the traditional PPN gauge, where $g_{ij}$  takes a diagonal form. After some algebraic manipulation, we obtain
\begin{align}\label{k}
    h^{(2)}_{00} &=\frac{(M_0Q_0 + 2M_1^2)}{2M_0Q_0 + 3M_1^2}\frac{\kappa U}{2\pi M_0},\\[1ex]
    h^{(2)}_{ij} &= \frac{M_0Q_0 + M_1^2}{2M_0Q_0 + 3M_1^2}\frac{\kappa U}{2\pi M_0}\delta_{ij},\\[1ex]
    \phi_2&=\frac{M_1}{2M_0Q_0+3M_1^2}\frac{\kappa U}{4\pi}.\label{k2}
\end{align}

To ensure that we correctly recover the Newtonian limit, we need to have the zero-zero component of the metric expressed as $g_{00}=-1 + 2U$ \cite{PoissonWill}, which in turns fixes the constant $\kappa$ appearing in (\ref{k}):
\begin{equation}\label{kk}
    \kappa=4\pi M_0 \left(\frac{2M_0Q_0 +M_1^2}{M_0Q_0 +2M_1^2}\right).
\end{equation}
Replacing (\ref{kk}) in Eqs. (\ref{k})--(\ref{k2}), helps us to determine the metric components and scalar field at second order: 
\begin{align}\label{so}
    h^{(2)}_{00} &=2U\\[1ex]
    h^{(2)}_{ij} &= \frac{2(M_0Q_0 + M_1^2)}{M_0Q_0 + 2M_1^2}\,U\delta_{ij},\\[1ex]
    \phi_2&=\frac{M_0M_1}{M_0Q_0+2M_1^2}\,U.
\end{align}
We notice that $h^{(2)}_{ij}$ and $\phi_2$ are influenced by the coupling functions $M$ and $Q$, which multiply the Ricci term and the kinetic term in the action, respectively. In contrast, the component $h^{(2)}_{00}$ is expected to remain unaffected by these coupling functions in order to recover the correct Newtonian limit.

We carry on by presenting the field equation in third order. In this case,  there is only one equation that needs to be solved, 
\begin{align}
    \nabla^2h^{(3)}_{0i} + \delta^{kl}\big(h^{(2)}_{kl,0i} -h^{(2)}_{ik,l0}-h^{(3)}_{0k,li}\big) \notag\\
    +\frac{2M_1}{M_0}\phi_2{}_{,0i} =\frac{2\kappa\rho v_i}{M_0}.
\end{align}
To solve the third-order PN equation, we have to use the already known second-order solutions and suppose that $h^{(3)}_{0i}$ is only proportional to the potentials $V_i$ and $W_i$, namely
\begin{align}
V_{i}&=\int\dfrac{\rho{}'v'_{i}d^{3}x' }{|\bs x- \bs x'|}\,,\\[1ex]
W_i  &= \int \frac{\rho' [{\bf v}' \cdot ({\bf x}-{\bf x}')](x-x')_i}
        {|{\bf x}-{\bf x}'|^3} \, d^3x'.
\end{align}
These potentials satisfy the following relations,
\begin{align}
    \nabla^2V_i&=-4\pi\rho v_i, \label{3a}\\
    \nabla^2\chi&=-2U, \label{3b} \\  
    \chi_{,0i}&=V_i-W_i, \label{3c} 
\end{align}
where $\chi$, the so-called superpotential, is an auxiliary function used to solve the third-order equation.

The set of equations (\ref{3a}--\ref{3c}) has as a formal solution the following metric component:
\begin{align}
    h^{(3)}_{0i}&=-\frac{(4+\lambda)M_0Q_0 + 2(3+\lambda)M_1^2}{2M_0Q_0 + 4M_1^2}\,V_i\notag\\
    &~- \frac{(4-\lambda)M_0Q_0 + 2(3-\lambda)M_1^2}{2M_0Q_0 + 4M_1^2}\,W_i,
\end{align}
where the constant $\lambda$ arises due to gauge freedom and will be fixed at the fourth order.

The zero-zero metric component at fourth order can be reformulated as follows:
\begin{widetext}
\begin{align}\label{fo}
16\pi^2M_0^2(2M_0Q_0+3M_1^2)^2\nabla^2h_{00}^{(4)} - 16\pi^2M_0M_1(2M_0Q_0+3M_1^2)^2\nabla^2\phi_4 +\varepsilon f_1M_1\kappa^2(2M_0Q_0+3M_1^2){\cal G}^{(4)}/8 \notag\\[1ex]
    + \kappa^2\vert\vec\nabla U\vert^2\big[(2M_0Q_0+3M_1^2)^2 +M_1^2(8M_1^2+4M_0Q_0-M_0M_2)\big] 
    =4\pi\kappa^2\rho UM_1^2(14M_1^2+8M_0Q_0-M_0M_2)\notag\\[1ex]
    - 16\pi^2 \kappa M_0(2M_0Q_0+3M_1^2)^2 (\rho\Pi + 3p + 2\rho v^2) 
    -  4\pi\kappa M_0M_1^2(2M_0Q_0+3M_1^2)U_{,00}.
\end{align}
\end{widetext}
In (\ref{fo}), ${\cal G}^{(4)}$ represents the fourth order approximation of the Gauss-Bonnet term, 
\begin{equation}
    {\cal G}^{(4)}=8U_{,ij}U^{,ij}-8(\nabla^2U)^2.
\end{equation}
It is important to note that terms such as $U_{,00}$ 
clearly manifest themselves as fourth-order in the PN expansion, taking into account that $\partial_{tt} \simeq  {\mathcal{O}}(v/c)^{2}$. A similar argument holds for the term $(\rho\Pi + 3p + 2\rho v^2)$, noticing that $p \simeq \rho \Pi \simeq \rho v^{2} \simeq {\mathcal{O}} (v/c)^{4}$ \cite{PoissonWill}.  Eq. (\ref{fo}) tells us that the term $\nabla^2h^{(4)}_{00}$ is coupled with the Laplacian scalar field $\nabla^2\phi_4$. Consequently, we must solve both the metric and the scalar-field Laplacian equations at the fourth order simultaneously. The latter equation is given by  
\begin{widetext}
\begin{align}\label{fo1}
-16\pi^2M_0(2M_0Q_0+3M_1^2)^3\nabla^2\phi_4
-\varepsilon f_1\kappa^2(2M_0Q_0+6M_0Q_0M_1^2+3M_1^4){\cal G}^{(4)}/8 \notag\\[1ex]
    + \kappa^2\vert\vec\nabla U\vert^2M_1^2\big[3M_1^3-M_0^2Q_1+M_0M_1(Q_0-3M_2)\big] 
    =16\pi^2 \kappa M_0M_1(2M_0Q_0+3M_1^2)^2 (\rho\Pi + 3p)\notag\\[1ex]
    4\pi\kappa^2\rho UM_1\big[6M_1^4+M_0M_1^2(8Q_0-3M_2)+2M_0^2(2Q_0^2-M_1Q_1+Q_0M_2)\big] 
    -  4\pi\kappa M_0M_1(2M_0Q_0+3M_1^2)^2U_{,00}.
\end{align}
\end{widetext}

Making use of the standard PPN potentials that are defined as
\begin{align}\label{pot4}
    &\nabla^2\Phi_1=-4\pi\rho v^2 &  \nabla^2\Phi_2=-4\pi\rho U\\[1ex]
    &\nabla^2\Phi_3=-4\pi\rho \Pi &  \nabla^2\Phi_4=-4\pi p,\\[1ex]
    &\chi_{,00} = {\cal A} + {\cal B} - \Phi_1,&
\end{align}
together with a new potential common to GB models,
\begin{equation}\label{potG}
    \nabla^2\Phi_{\cal G}=-4\pi{\cal G}^{(4)},
\end{equation}
the solutions are found to be
\begin{align}
h_{00}^{(4)}=& - \kappa^2 Z_{1} U^{2}+ \kappa Z_{2}\Phi_{1}{} + \kappa^2 Z_{3} \Phi_{2}{}+ \kappa Z_{4}\Phi_{3}{} \\ \nonumber
&+\kappa Z_{5}\Phi_{4}{} +\epsilon\kappa^2 Z_{6} \Phi_{\mathcal{G}}{},\nonumber \\[2ex]
\phi_{4}=&  ~\kappa^2 Y_{1} U^{2}+ \kappa  Y_{2}\Phi_{1}{}+\kappa^2 Y_{3}\Phi_{2}{}+ \kappa  Y_{4}\Phi_{3}{}\nonumber\\
&+\kappa Y_{5}\Phi_{4}{}+ \kappa  Y_{6}\mathcal{A}+
\kappa  Y_{7}\mathcal{B}  + \varepsilon \kappa^2 Y_{8} \Phi_{\mathcal{G}},
\end{align}
where coefficients $Y_{i}$ and $Z_{i}$  are listed in the Appendix A, provided these ones are very lengthy expressions.  It is important to emphasize that we double checked our expressions using the xPPN package developed in Mathematica by Hohmann \cite{Hohmann:2020muq}.  The formal solutions of the fourth-order PN potentials involving
are given in terms of the Poisson integral as follows \cite{PoissonWill},
\begin{align}
\Phi_1 &=\int\dfrac{\rho{}' v'^{2}}{|\bs x- \bs x'|}\,d^{3}x',\quad \Phi_2 =\int\dfrac{\rho{}' U'}{|\bs x- \bs x'|}\,d^{3}x',\\[1ex]
\Phi_3 &=\int\dfrac{\rho{}' \Pi'}{|\bs x- \bs x'|}\,d^{3}x',\quad \Phi_4 =\int\dfrac{p'}{|\bs x- \bs x'|}\,d^{3}x',\\[1ex]
\PhiG&=\!\int\dfrac{\G}{|\bs x- \bs x'|}\,d^{3}x'\!, \ {\cal A} = 
        \int \frac{\rho'{\bf v}'\!\cdot\!({\bf x}-{\bf x}')}
        {|{\bf x}-{\bf x}'|^3}\,d^3x'\!,
\\[1ex]
        & \qquad\qquad {\cal B}=\int\frac{\rho'({\bf x}-{\bf x}')}{|{\bf x}-{\bf x}'|}\cdot\frac{d\vec{v'}}{dt}\,d^3x'.
\end{align}
 In the next section, we are going to focus on the observational predictions coming from different scalarized theories; in particular, we will look at how the parameter space can be restricted by using the latest bounds on the PPN parameters.

\section{The Extended PPN approach}
Having completed the PPN analysis with a fourth-order expansion, we are now poised to explore a range of theoretical and observational implications. To begin, it is essential to highlight several significant findings that have emerged from this work. First, we observe the emergence of the nonstandard potential $\Phi_{\cal G}$ in the fourth-order metric component, which, as anticipated, is absent in the PPN framework. Under this configuration, it is futile to extract the PPN parameters without a thorough understanding of how the new potential interacts with the others (see Refs. \cite{Toniato:2019rrd,Toniato:2021vmt} for details).

In particular, the significance of $\Phi_{\cal G}$ has been illustrated in previous work, demonstrating its critical role in influencing the periastron precession \cite{Toniato:2024gtx}. In other words, the condition that the PPN parameter $\beta$ equals 1 is insufficient to ensure consistency with observations of Mercury's perihelion advance. This aspect is not exclusive to the theories here considered. In fact, the same effect on the $\beta$ parameter is also found in massive Brans-Dicke models \cite{alves2024}.

Within the context of extended PPN parameters, the  metric that incorporates the new effects associated with $\Phi_{\cal G}$ is expressed as follows:
\begin{align}
    g_{00}=&-1 +2U - 2\beta_1 U^2  +2\beta_2\Phi_{\cal G} - 2\xi\Phi_W  \notag\\
    &+(2\gamma +2+\alpha_3+\zeta_1-2\xi)\Phi_1\notag\\
    &+2(2\gamma - 2\beta +1+\zeta_2+\xi)\Phi_2 + 2(1+\zeta_3)\Phi_3 \notag\\
    &+2(3\gamma +3\zeta_4-2\xi)\Phi_4 -(\zeta_1-2\xi){\cal A},\label{metricPPN00}\\[2ex]
    g_{0i}=& -\frac{1}{2}(3+\alpha_1-\alpha_2+4\gamma-2\xi+\zeta_1)V_i  \notag\\
    &\quad -\frac{1}{2}(1+\alpha_2+2\xi-\zeta_1)W_i,\label{eppn}\\[2ex]
    g_{ij}=& \ (1+2\gamma U)\delta_{ij}.\label{metricPPNij}
\end{align}
The ansatz above recovers the ordinary PPN framework if $\beta_2=0$ and $\beta_1=\beta$. As demonstrated in Ref. \cite{Toniato:2024gtx}, the physical interpretation of all standard PPN parameters, except $\beta$, remains consistent. In this extended PPN framework, both $\beta_1$ and $\beta_2$ play critical roles in the periastron precession of a binary system. The rate of secular advancement can be written as,
\begin{align}
    \dot{\tilde\omega} =& ~ 3\left(\frac{2\pi}{P}\right)^{5/3}\frac{m^{2/3}}{1-e^2}\bigg[\frac{1}{3}(2+2\gamma-\beta_1) \notag\\
    &\qquad +\frac{\mu}{6m}(2\alpha_1-\alpha_2+\alpha_3+2\zeta_2) + \frac{J_2R^2}{2mp}\bigg] \notag\\
    &~ + 6\beta_2\left(\frac{2\pi}{P}\right)^{3}\frac{4+e^2}{(1-e^2)^3}.
    \label{dot-omega}
\end{align}
Here, $P$ represents the orbital period, as derived from Kepler's third law, given by the relation $P^2=4\pi^2p^3/m(1-e^2)^3$, where $m$ denotes the total mass of the system and $\mu$ is the reduced mass. Meanwhile, $p$ and $e$ refer to the semilatus rectum and the eccentricity of the elliptical orbit, respectively. The factor $J_{2}$ encapsulates the effects of the quadrupole moment generated by the dominant body, with $R$ indicating its radius. In conservative theories, it is found that all the parameters $\alpha$ and $\zeta$ are equal to zero. Furthermore, since light propagation tests fully constrain the $\gamma$ parameter, the only relevant parameters for the periastron shift are the $\beta$ parameters.

After comparing the extended PPN metric \eqref{eppn} with the solutions obtained for the scalar-EGB theory, one can now extract the corresponding PPN parameters:
\begin{align}\label{ppns}
    \beta_1&=1+\frac{M_0M_1^2(Q_0M_1^2 -2M_0Q_0M_2 +M_0Q_1M_1)}{4(M_0Q_0+2M_1^2)^2(2M_0Q_0 +3M_1^2)}\,, \\
    \beta_2&=\frac{3\varepsilon f_1M_1(M_0Q_0+M_1^2)}{32\pi(M_0Q_0+2M_1^2)(2M_0Q_0 +3M_1^2)},\label{beta2}\\[2ex]
    \gamma&=\frac{1}{2}\left(1+\frac{M_0Q_0}{M_0Q_0+2M_1^2}\right)\label{gamma}\\[2ex]
\alpha_1&=\alpha_2=\alpha_3=\xi=\zeta_1=\zeta_2=\zeta_3=\zeta_4=0.
\end{align}
With the parameters $\xi$, $\alpha$, and $\zeta$ set to zero, the scalar EGB models under consideration are entirely conservative theories. This ensures that there are no energy and momentum conservation violations, nor are there any violations of the equivalence principles \cite{Will:1993ns}. Compared to standard General Relativity, only the parameters $\beta$ and $\gamma$ can exhibit variations, depending on the specific model. Each model is characterized by a specific selection of the functions $M(\Phi),~Q(\Phi)$ and $f(\Phi)$, which serve as different characteristics for the scalarization process. As a general rule, it can be noted that all the scalar-EGB theories described in Sec. \ref{sec:scalarEGB} exhibit an influence on both the perihelion shift and the results of light motion tests.

In relation to the parameter $\gamma$ (\ref{gamma}), we find that the principal modifications, compared to General Relativity, arise from the coupling function that multiplies the Ricci term, as well as the factor present in the kinetic term of the action (\ref{ac1}). This observation aligns with the alterations introduced in Brans-Dicke theory or in conformal gravity, where a term such as $\xi \Phi^{2}R$ serves as a pertinent example, or in the case of the Ricci-EGB model with $M=1-(\beta/4) \Phi^2$. A similar scenario occurs with the parameter $\beta_1$. However, the context changes when we consider the $\beta_2$ parameter. As indicated in Eq. (\ref{beta2}), this parameter is also affected by its coupling with the Gauss-Bonnet term.  For example, the Ricci-EGB model faces a direct constraint arising from the parameter $\beta_2$, which is derived from observations of the rate of change $\dot{\omega}$ in binary systems.
Although the scalarized EGB model includes a Gauss-Bonnet term in its action, direct computations reveal that $\beta_{2}=0$. This indicates that the model successfully satisfies all constraints of the solar system. On the other hand, the equivalent DEF model faces comparable constraints, although this time they stem from the parameter $\beta_1$ in the rate of secular advancement (\ref{dot-omega}), and not through $\beta_{2}$ provided $f(\Phi)=0$. To enhance our understanding of the parameter space associated with each theory, derived from the extended PPN approach, we present the Table \ref{tab:parameters} which shows the specific values.
\begin{table*}[t!]
\renewcommand{\arraystretch}{2}
    \centering
        \caption{The (extended) PPN parameters that can differ from those in General Relativity across various scalar-EGB models.}\label{tab:parameters}
    \begin{tabular}{c|c|c|c|c}
    \hline\hline
        \textbf{Model} & ~$\boldsymbol{M(\Phi)}$~ & ~$\boldsymbol{Q(\Phi)}$~ & ~$\boldsymbol{f(\Phi)}$~ & \textbf{PPN parameters} \\
        \hline
        Brans-Dicke & $\Phi$ & $\dfrac{\omega}{\Phi}$ & 0 & $\beta_1=1,~\beta_2=0,~\gamma=\dfrac{1+\omega}{2+\omega}$\\[1ex]
        \hline
        Scalarized EGB & 1 & 1 & free & $\beta_1=1,~\beta_2=0,~\gamma=1$\\
        \hline
        \multirow[c]{3}{*}{Ricci EGB} & \multirow[c]{3}{*}{$1-\dfrac{\lambda}{2}\,\Phi^2$} & \multirow[c]{3}{*}{1} & \multirow[c]{3}{*}{$\frac{\sigma}{2}\,\Phi$} & $\beta_1=1+\frac{\lambda^2\phi_0^2(2-\lambda\phi_0^2)}{[2+\lambda(4\lambda-1)\phi_0^2]^2[2+\lambda(3\lambda-1)\phi_0^2]}$\\ & & & & $\beta_2=\frac{-3\lambda\sigma\varepsilon\phi_0^2[2+\lambda(2\lambda-1)\phi_0^2]}{64\pi[2+\lambda(4\lambda-1)\phi_0^2][2+\lambda(3\lambda-1)\phi_0^2]}$ \\ & & & &  $\gamma=\frac{1}{2}\left[1+\frac{2-\lambda\phi_0^2}{2+\lambda(4\lambda-1)\phi_0^2}\right]$\\[1ex]
        \hline
        \multirow[c]{3}{*}{Equivalent DEF} & \multirow[c]{3}{*}{$1-\dfrac{\lambda}{4}\,\Phi^2$} & \multirow[c]{3}{*}{1} & \multirow[c]{3}{*}{0} & $\beta_1=1+\frac{8\lambda^2\phi_0^2(4-\lambda\phi_0^2)}{[8+\lambda(4\lambda-2)\phi_0^2]^2[8+\lambda(3\lambda-2)\phi_0^2]}$\\ & & & & $\beta_2=0$ \\ & & & &  $\gamma=\frac{1}{2}\left[1+\frac{8-2\lambda\phi_0^2}{8+\lambda(4\lambda-2)\phi_0^2}\right]$\\[1ex]
        \hline\hline
    \end{tabular}
\end{table*}

\section{Astrophysical Constraints}

\subsection{Solar-System}
\begin{figure}[t]
\includegraphics[width=\columnwidth]{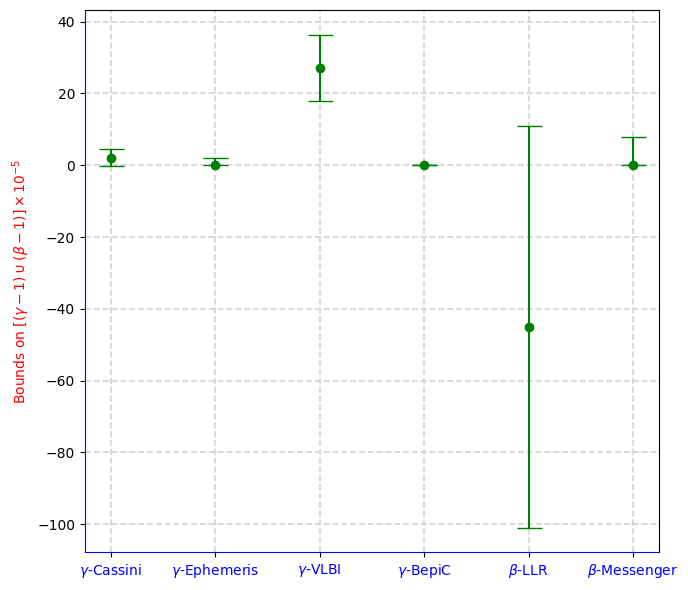}
\caption{Bounds on $\gamma$ and $\beta$ parameters from different analysis: Cassini mission \cite{Bertotti:2003rm}, Lunar Laser Ranging data (LLR) \cite{Hofmann:2018myc}, planetary ephemeris \cite{Mariani:2023rca}, geodetic VLBI technique \cite{2018A&A...618A...8T}, BepiColombo mission (estimated) \cite{DeMarchi:2019lei} and MESSENGER \cite{Park:2017zgd}. The LLR constraint is of the same order of magnitude than Cassini but forcing $\gamma$ to assume values only smaller than 1. It is noteworthy that the BepiColombo mission is set to probe one order of magnitude below the existing bounds, specifically $\mathcal{O}(10^{-6})$. Consequently, the associated error bar becomes imperceptible on the scale of $\mathcal{O}(10^{-5})$.}
\label{fig:GammaBetaBounds} 
\end{figure}

The most accurate constraints derived from the Solar System for the PPN parameters $\gamma$ and $\beta$ come from measurements of the gravitational delay of electromagnetic waves travel time, by the Cassini spacecraft \cite{Bertotti:2003rm}, and Mercury's perihelion advance rate, which is given by MESSENGER mission \cite{Park:2017zgd} (see Fig. \ref{fig:GammaBetaBounds}).
Once the EPPN framework adopted before does not change the equations of motion of photons, the Cassini stringent bound can be directly applied  to the parameter $\gamma$, which is expressed as 
\begin{equation}
    \gamma_{\rm{cassini}}-1=(2.1\pm 2.3)\times 10^{-5}.\label{cassini}
\end{equation}
Conversely, with the EPPN modification $\beta\rightarrow (\beta_1,\,\beta_2)$, the constraint on Mercury's perihelion shift must be applied through Eq. \eqref{dot-omega}, where 
\begin{equation}\label{messenger}
    \dot{\tilde{\omega}}=(42.9799\pm 0.0009)\,{\rm arcsec/century}.
\end{equation}
In cases where $\beta_2=0$, one can apply MESSENGER's constraint for the PPN parameter $\beta$ direct on $\beta_1$, resulting in $\beta_1-1=(-2.7\pm 3.9)\times 10^{-5}$. In the following, we will explore the implications of the above constraints upon the models presented in Table \ref{tab:parameters}.

Once the generic scalar-EGB model considered includes also the ordinary Brans-Dicke (BD) theory for $M=\Phi,~Q=\omega/\Phi$ and $f=0$, we added the resulting EPPN parameters for the sake of completeness. As it is well known, in this context, the only parameter that deviates from General Relativity is $\gamma$. This results in a high value for the BD parameter, accordingly to the Cassini bound, namely $\omega>4.35\times 10^{4}$. For the scalarized EGB models, Table \ref{tab:parameters} shows that they are indistinguishable from General Relativity in light of the constraints derived from solar system observations. However, other Scalarized models, such as the Ricci EGB and the equivalent DEF models, are indeed influenced by these solar system constraints.

Now we turn our attention to the Equivalent-DEF theories, which are models covered by the standard PPN formalism, since $\beta_2$ is identically null. There is a particular case of interest that is to assume small values of $\phi_0$, which gives the relation $\lambda\phi_0\lesssim10^{-2}$, after considering Mercury's perihelion shift constraint. Thus, the expectation value of the scalar field must be smaller as its coupling to the gravitational sector becomes stronger. In the limit where $\lambda$ is small, we recover the Scalarized-EGB models and the agreement with Solar System tests.

\begin{figure}[t]
\includegraphics[width=\linewidth]{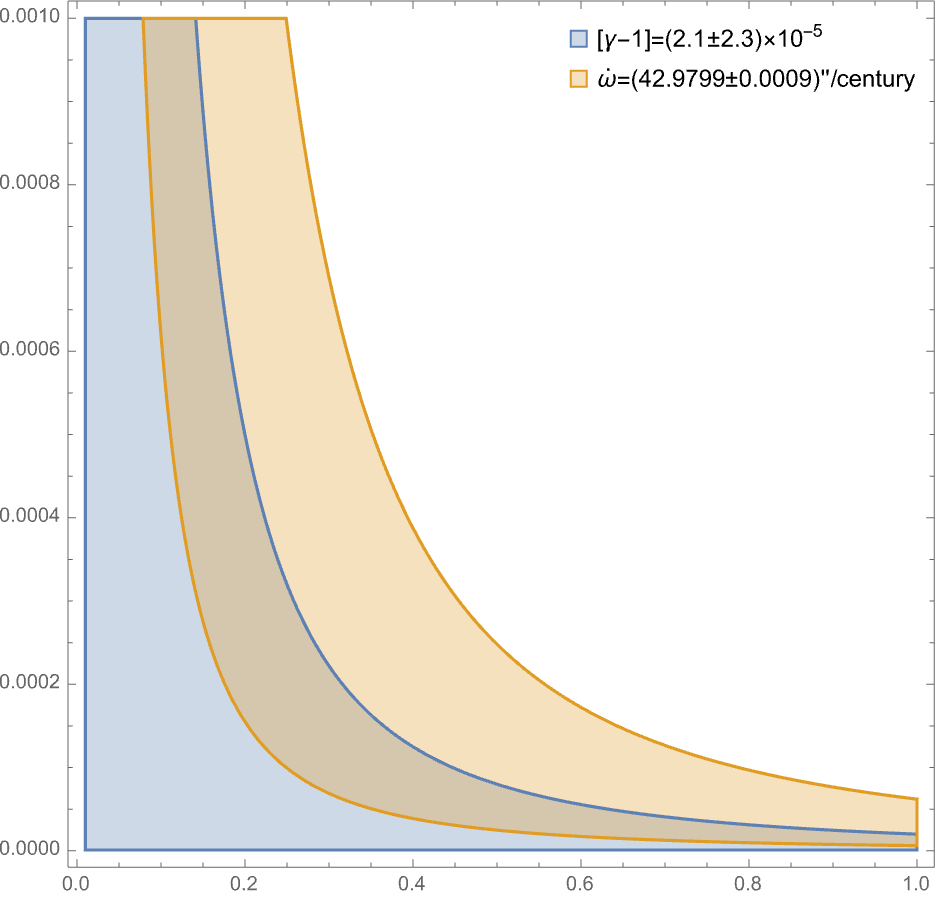}
\caption{We present the joint constraints on the $\gamma$ parameter derived from the Cassini mission alongside the MESSENGER observations related to the precession rate of Mercury's perihelion. This analysis focuses on the Ricci-EGB model, where we define $x=\lambda$  within the range $ [0,1]$ and $y=\phi^{2}_{0}$ constrained to the interval $[0, 0.001]$. 
\label{fig:CassiniMessenger}} 
\end{figure}

When considering the Ricci EGB models, an alternative approach can be employed in this analysis. For example, we could integrate the constraints of each theory within the $\lambda$--$\phi^{2}_{0}$ plane. By combining the bounds from Cassini [cf. \eqref{cassini}] with the measurements from the MESSENGER mission [cf. \eqref{messenger}], we delineate the region in parameter space where both missions converge, as can be seen in Fig. \ref{fig:CassiniMessenger}. It is noteworthy that the constraint from Cassini provides a significantly tighter bound compared to the MESSENGER limit on $\dot{\tilde{\omega}}$. Importantly, the value of $\lambda$ is not required to be small or positive; however, our analysis indicates that the squared expectation value, $\phi^{2}_{0}$, remains restricted to very small values.

\subsection{Strong-Lensing System}

An attractive constraint on $\gamma$ arises from the study of strong gravitational lensing (SGL) in the context of galaxies. By operating within a weak-field approximation across kiloparsec scales, a galaxy can be conceptualized as a lens that generates multiple images of a background source. In 2006, Bolton \emph{et al.}  utilized the SLACS survey, analyzing 15 elliptical galaxies that exhibited gravitational lensing \cite{Bolton:2006yz}. Their findings allowed them to constrain the PPN parameter, producing a measurement of $\gamma_{\rm{Bolton}-2006}=0.98\pm 0.07$.  Subsequently, Schwab \emph{et al.} expanded upon this work by analyzing 53 galaxy-scale strong lenses from the same SLACS survey\cite{2010ApJ...708..750S}. Their refined results yielded a measurement of $\gamma_{\rm{Schwab}-2009}=0.98\pm 0.07$, improving the precision of previous findings. These seminal works showed the utility of strong gravitational lensing as an effective method for investigating cosmological models and gravitational theories.  

\begin{figure}[t]
\includegraphics[width=\columnwidth]{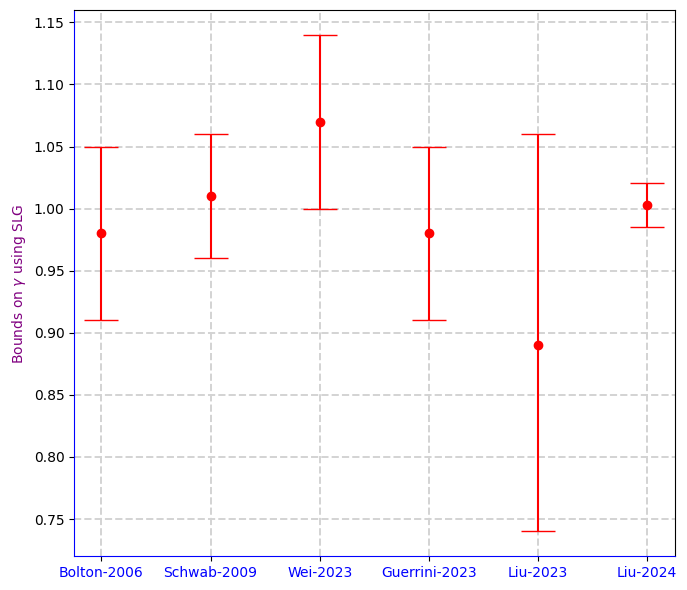}
\caption{Bounds on $\gamma$ obtained by different strong lensing systems. 
\label{fig:2}} 
\end{figure}

By combining Pantheon supernova Ia observations within the framework of a flat universe to compute cosmological distances, alongside several strong gravitational lensing systems that produce the distinctive Einstein rings,  Wei \emph{et al.} obtained $\gamma_{\rm{Wei}-2023}=1.07\pm 0.07$. However, their analysis revealed a significant degeneracy with respect to the non-zero curvature parameter $\Omega_{k}$.
Guerrini and M\"ortsell investigated the implications of a fifth force on a sample of 130 early-type galaxies with E/S0 morphologies, drawn from multiple surveys, including SLACS, BELLS, and BELLS GALLERY \cite{Guerrini:2023pre}. Their analysis led to a measurement of $\gamma_{\rm{Guerrini}-2023}=1.07\pm 0.07$. However, it is important to note that this value is considerably reduced when considering the effects of a screening scale characterized by a Compton length of $\lambda_{g}=0.2$ Mpc. A summary of the different bounds on $\gamma$ is shown in Fig. \ref{fig:2}.

In 2023, Liu \emph{et al.} pursued a promising approach by leveraging reconstructed unanchored distances in conjunction with data sets from the four H0LiCOW lenses \cite{Liu:2023ulr}. In particular, their analysis yielded a measurement of $\gamma_{\rm{Liu}-2023}=0.89^{+0.17}_{-0.15}$, alongside an estimation of the Hubble constant at $H_{0}=72.9^{+2.0}_{-2.3}  \rm{km}^{-1}  \rm{s}^{-1} \rm{Mpc}^{-1}$. Subsequently, Liu et al. integrated strongly lensed systems with gravitational wave signals from coalescing neutron stars to mitigate various astrophysical biases and degeneracies \cite{Liu:2024bre}. Their comprehensive analysis resulted in a refined measurement of the parameter, which yielded $\gamma_{\rm{Liu}-2024}=1.003\pm 0.018$.

In this scenario, the angular separations between these images approach the well-known Einstein radius $\theta_{E}$ \cite{Wei:2022rcb}:  
\begin{align}
    \theta_{E}=\sqrt{\frac{\gamma_{\rm{sgl}} +1}{2}} \Big(\frac{4GM_{E}}{c^2}\frac{D_{ls}}{D_{s}D_{l}}\Big)^{1/2}.
\end{align}
Here, $M_{E}$ denotes the gravitational mass associated with the lensing galaxy, while $D_s$ represents the distance of the angular diameter to the source. Furthermore, $D_l$ indicates the distance of angular diameter from the lens itself and $D_{ls}$ refers to the distance of angular diameter that separates the lens from the source.  When examining the lensing effects produced by a galaxy located at a cosmological distance of $D = 1{\rm Gpc}$ with a mass $M =10^11{M}_{\odot}$, we find that the corresponding Einstein radii are given by
\begin{align}
    \theta_{E}=(0''.9)\Big(\frac{M}{10^{11}M_{\odot}}\Big)^{1/2}\Big(\frac{D}{\rm{Gpc}}\Big)^{-1/2}.
\end{align}

In the context of SGL,  the metric element is expressed as $ds^2=a^{2}(\tau)[-(1+2\Psi_{N})d\tau^2+(1+2\gamma \Phi_{N})dx^{i}dx_{i}]$, where $\tau$ is the conformal time, and $a(\tau)$ is the scale factor,  which accounts for the expansion of the universe at high redshift. This formulation captures the essential dynamics of gravitational lensing in a cosmological setting, incorporating both the effects of scalar potentials and the underlying expansion of the universe. The adopted notation is $\Psi_{N}=U$ and $\Phi_{N}=U\gamma$, which in turn leads to a straightforward relationship, allowing us to express the slip as $\eta_{\rm{\textbf{PPN}}}=\Phi_{N}/\Psi_{N}=\gamma$.

In general, the density profile employed for galaxies, which serve as gravitational lenses, can be represented by power laws of the form $\rho(r) \propto r^{-\alpha}$ and $\nu(r) \propto r^{-\delta}$. An anisotropic factor, $\beta = 1 - \sigma^{2}_{t}/\sigma^{2}_{r}$, is generally assumed to be constant \cite{2006EAS....20..161K}. With these profiles established, the radial Jeans equation can be utilized to derive the radial velocity dispersion of stars within the galaxy, denoted as $\sigma_{r}(r)$, from which the total mass, $M(r)$, can be inferred. In practice, the total mass can be expressed in terms of the Einstein ring radius, $R_{E} = \theta_{E} D_{l}$ and the dynamical mass, $M^{\rm{dyn}}_{E} = M^{\rm{grav}}_{E}$, leading to the relationship:
\begin{align}
    \sigma^{2}_{r}= \mathbf{J} \frac{GM^{\rm{dyn}}_{E}}{R_{E}} \left(\frac{r}{R_{E}}\right)^{2-\alpha},
\end{align}
where $\mathbf{J}$ encompasses various numerical factors and parameters, as discussed in \cite{2006EAS....20..161K}.
To align with observational data, it is imperative to utilize the projected version of the line-of-sight (LOS) velocity dispersion, $\sigma^{2}_{LOS}(R)$. In this context, the coordinates are related through the equation $r^2=R^2+Z^2$, where $R$ represents the projected radial coordinate and $Z$ denotes the coordinate along the LOS. Furthermore, we incorporated an aperture weighting function, formulated as $\mathcal{A}(R)\simeq e^{-R^{2}/2D^{2}_{l}\hat{\sigma}_{\rm{atm}}}$  \cite{2010ApJ...708..750S}. The expression for the weighted projected LOS velocity dispersion, $\langle\sigma^{2}_{LOS,\star}(R)\rangle$, has been discussed in various sources (see, for instance, Refs. \cite{2010ApJ...708..750S,Wei:2022rcb}). The rescaled seeing factor is given by  
\begin{align}
\hat{\sigma}^2_{\rm{atm}}=\sigma^2_{\rm{atm}}\Big[1+\frac{1}{4}\big(\frac{\theta_{\rm{ap}}}{\sigma_{\rm{atm}}}\big)^2+ \frac{1}{40}\big(\frac{\theta_{\rm{ap}}}{\sigma_{\rm{atm}}}\big)^4\Big].
\end{align}
Here, $\sigma_{\rm{atm}}$ represents the seeing factor, which varies according to the specific survey, such as SLACS, BELLS, SDSS, and S4TM-but generally falls within the range of 0.3 to 1.8 arcsec. In addition, $\theta_{\rm{ap}}$ denotes the radius of the fiber aperture. 
\begin{figure}[t]
\includegraphics[width=\columnwidth]{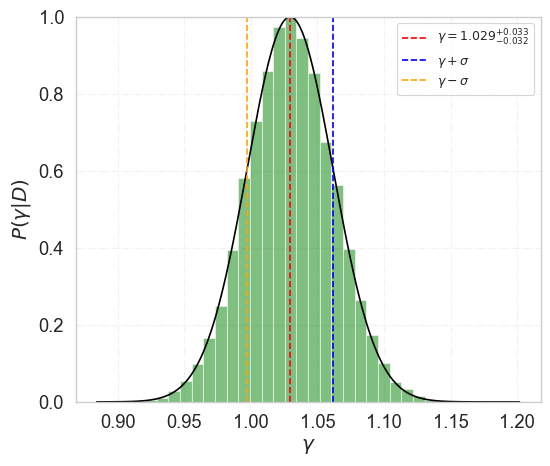}
\caption{Bounds on $\gamma$ derived from a MCMC simulation based on 100 strong lensed galaxies.
\label{fig:3}} 
\end{figure}

In what follows, we focus on a sample of 100 early-type lens galaxies reported in \cite{Chen:2018jcf} provided that such a catalog represents one of the most extensive collections encompassing both high-resolution imaging and stellar dynamical data. We impose a limit on the sample primarily due to certain concerns associated with the seeing factor. More importantly, our intention is to illustrate how the strong lensing data can be used to derive alternative bounds on $\gamma$, distinct from those obtained in the strong field regime or restricted to our local solar system measurements.  We performed a global Monte Carlo simulation within a four-dimensional parameter space, characterized by $N_{\rm{burn}}=1.000$ and $N_{\rm{steps}}=20.000$, and tested
the convergence of the MCMC chains by using the Gelman-Rubin criterion $|R-1| < 0.01$. Later, we marginalized over all parameters except for $\gamma$, and the posterior was expressed as a Gaussian distribution, $P(\gamma|\rm{data})\propto e^{-(\gamma-\gamma_c)^{2}/\sigma^{2}_{\gamma_c}}$, as shown in Fig. \ref{fig:3}. The best-fit value obtained is $\gamma=1.029^{+0.033}_{-0.032}$, which shows strong consistency with the measurement $\gamma_{\rm{Wei}-2023}=1.07\pm 0.07$, along with other similar estimates.

At this point, a pertinent question arises concerning the implications for scalarized models. To illustrate this, we examine the Ricci-EGB scenario, specifically selecting the case where $\lambda=1/4$ for the sake of brevity.  Inserting the bound $\gamma=1.029^{+0.033}_{-0.032}$ into the formal expression of $\phi^{2}_{0,-}=16(1-\gamma_{-})$, we obtained a low value for $\phi^{2}_{0,-}\simeq 6/125$. Notice that when employing four H0LiCOW lenses, we attain a higher value for the Ricci-EGB scenario; that is, $\phi^{2}_{0,\rm{HoliCow}}\simeq  104/25$ \cite{Liu:2023ulr}.

We conclude this section by examining the equivalent DEF model alongside the Brans-Dicke theory.\footnote{It is noteworthy that the scalarized EGB model does not necessitate further analysis, as it yields $\gamma=1$. In fact, the equivalent DEF model leads to the same finding once we choose $\lambda=1/4$.}  Within the framework of the Brans-Dicke theory, the constraints obtained from the estimation of $\gamma$ using SGL data establish bounds of $-18<\omega<331$. This range notably diverges from the limits reported in local constraints, as emphasized in previous research \cite{Bertotti:2003rm,Mariani:2023rca}. Nevertheless, this theoretical framework offers a compelling approach for evaluating parameter bounds, particularly in systems characterized by pronounced gravitational lensing effects. Looking ahead, it would be worthwhile to revisit the aforementioned analysis by integrating gravitational wave signals from coalescing neutron stars. This approach could help address various astrophysical biases and degeneracies, as discussed in \cite{Liu:2024bre}. Such an incorporation could enhance the robustness of our findings and provide deeper insights into the underlying physics.

\section{Summary}

We undertook a comprehensive exploration of a class of scalarized theories that exhibit GB coupling, employing the PPN method as an effective tool to delineate the distinctive characteristics of this class.  Through this approach, we systematically compared these theories in terms of their modifications to GR.  Our primary focus was on the new characteristics and modifications introduced in the parameters of PPN $\gamma$ and $\beta$. Given that the scalarized EGB model is situated within a conservative metric theory framework, we did not anticipate any violations of the equivalence principle or momentum conservation. Nevertheless, the coupling functions $M(\Phi),~Q(\Phi)$ and $f(\Phi)$, which reveal distinct properties during the scalarization process in black hole and neutron star systems, have been linked to several deviations from General Relativity in the weak-field regime, as evidenced by multiple missions. These deviations are not limited to weak-field scenarios; they also manifest in strong-field conditions, highlighting the broader implications of these coupling functions across different astrophysical environments. 

One compelling aspect of the PPN method is its ability to differentiate between various, albeit similar, theories that may not be on the same footing. From this perspective, we found that scalarized EGB models require an extended formulation of the PPN parameters. This extension is particularly significant, as it reveals that the parameter $\beta$ exhibits distinct effects when analyzing different scenarios. To be more precise, the Ricci-EGB model encounters direct constraints stemming from the parameters $\beta_2$ and $\beta_1$, which are derived from observations of the rate of change $\dot{\tilde\omega}$ in binary systems, assuming that both parameters are non-zero. In contrast, the Equivalent-DEF model uniquely features $\beta_{2}=0$, making it consistent with the standard PPN approach and their derived bounds. Moreover, the Scalarized-EGB model has the same PPN parameters as GR, making it indistinguishable from the latter regarding Solar System tests [cf. Table \ref{tab:parameters} and Eq. (\ref{dot-omega})].
These attributes highlight a fundamental distinction in how all these models respond to observational data and the implications for their respective gravitational theories.

With respect to observational bounds, we primarily focused on extracting constraints to the model parameters utilizing the current most accurate data for the Shapiro time delay and Mercury's perihelion shift. They are extracted from the Cassini mission, which reports $[\gamma_{\rm{cassini}}-1]=(2.1\pm2.3)\times10^{-5}$\cite{Bertotti:2003rm}, and from the MESSENGER measurements, yielding $\dot{\tilde\omega}\leq (42.9799 \pm 0.0009)\, \rm{arcsec/century}$ \cite{Park:2017zgd}. For the Equivalent-DEF model, we obtain that both constraints are satisfied provided that holds the relation $\lambda\phi_0\lesssim10^{-2}$.
In the framework of the Ricci-EGB model, we have integrated both constraints. This approach yields the allowed regions in the $\lambda$--$\phi^{2}_{0}$ parameter plane [cf. Fig. \ref{fig:CassiniMessenger}]. In particular, both constraints exhibit an intersection region at the $1\sigma$ level. Our analysis indicates that while the value of $\lambda$ is not restricted to be small or positive, the squared expectation value $\phi^{2}_{0}$ must remain very small.

Finally, we constrained the $\gamma$ PPN parameter by analyzing a catalog of 100 early-type lens galaxies as reported in \cite{Chen:2018jcf} (see  Fig. \ref{fig:2}).  We obtained the best fit value of $\gamma=1.029^{+0.033}_{-0.032}$, showing strong consistency with the measurement $\gamma_{\rm{Wei}-2023}=1.07\pm 0.07$, along with other similar estimates. In the Ricci-EGB scenario with $\lambda=1/4$, the SGL bound on $\gamma$ implied that the squared value of the scalar field in vacuum must be low, $\phi^{2}_{0}<10^{-2}$. The latter value would increase by two orders of magnitude if the four H0LiCOW lenses were included in the analysis without bias, as indicated in \cite{Liu:2023ulr}. To further enhance our current analysis, it is essential to investigate other systems that exhibit fewer astrophysical biases and degeneracies.

\section{Acknowledgments}
We acknowledge the use of the \textbf{xPPN}-package \url{https://github.com/xenos1984/xPPN} \cite{Hohmann:2020muq} as an inspiration to double-check our theoretical calculations. We used several packages such as \textbf{Astropy} \cite{Astropy:2013muo}, \textbf{Numpy} \cite{vanderWalt:2011bqk}, \textbf{Matplotlib} \cite{Hunter:2007ouj}, \textbf{Scipy} \cite{Virtanen:2019joe}, \textbf{Pandas} \cite{ mckinney-proc-scipy-2010}, \textbf{Seaborn}\cite{Waskom2021}, \textbf{Emcee} \cite{2013PASP..125..306F}.  M.G.R was partially supported by W.F.S.A. and J.D.T. is grateful for the partial support of FAPES, grant No. 1020/2022. 

\section{Appendix A:~ Coefficients of $h^{(4)}_{00}$ and $\phi^{(4)}$}
Here we list the main coefficients of $h^{(4)}_{00}$ and $\phi^{(4)}$:
\begin{widetext}
\begin{align}
Z_{1}&=\frac{M_1^2\big[8 M_0^3M_1 + 81 M_0 Q_0 M_1^2 + 48 M_1^4 + M_0^2 (44 M_1^2 + M_0 Q_1 - 2 Q_0 M_2)\big]}{32 \pi^2 M_0^2 (2 M_0 Q_0 + 3 M_1^2)^3},\\
Z_{2}&=\frac{1 }{2 \pi M_0},\\
Z_{3}&=\frac{\big[8 M_0^3 M_1^3 + 37 M_0 Q_0 M_1^4 + 12 M_1^6 + M_0^2 M_1^2 (32 M_1^2 -  M_0 Q_1 + 2 Q_0 M_2)\big]}{16 \pi^2M_0^2 (2 M_{0}Q_0 + 3 M_1^2)^3}, \\ 
Z_{4}&=\frac{(M_0 Q_0 + 2 M_1^2)}{2\pi M_0(2M_0Q_0 + 3M_1^2)},\\
Z_{5}&= \frac{3(M_0 Q_0 + M_1^2)}{2\pi M_0(2M_0Q_0 + 3M_1^2)},\\
Z_{6}&= \frac{3 f_1 M_1^3(M_0^2 + 3 M_0 Q_0 + 2 M_1^2)}{256 \pi^3 M_0^2 (2 M_0 Q_0 + 3 M_1^2)^3},\\
Y_{1}&= \frac{M_1^2\big[3 M_1^3 -  M_0^2 Q_1 + M_0 M_1(Q_0 - 3 M_2)\big]}{32 \pi^2 M_0 (2 M_0 Q_0 + 3 M_1^2)^3},\\
Y_{2}&=\frac{M_1}{8\pi(2M_0 Q_0 + 3M_1^2)},\\
Y_{3}&=\frac{M_1\big[7 M_0 Q_0 M_1^2 + 3 M_1^4 + M_0^2 (4 Q_0^2 -  M_1 Q_1 + 2 Q_0 M_2)\big] }{16 \pi^2 M_0 (2 M_0 Q_0 + 3 M_1^2)^3},\\
Y_{4}&=\frac{M_1}{4\pi(2M_0 Q_0 + 3M_1^2)},\\
Y_{5}&= -\frac{2M_1}{4\pi(2M_0 Q_0 + 3M_1^2)},\\
Y_{6}&=-\frac{ M_1}{8\pi(2M_0 Q_0 + 3M_1^2)},\\
Y_{7}&=-\frac{ M_1}{8\pi(2M_0 Q_0 + 3M_1^2)},\\
Y_{8}&= \frac{f_1 (2 M_0^2 Q_0^2 + 6 M_0 Q_0 M_1^2 + 3 M_1^4)}{512 \pi^3 M_0 (2 M_0 Q_0 + 3 M_1^2)^3}.
\end{align}
\end{widetext}

\section{Appendix B:~The PPN  parameters in GR}\label{appendix}
The arrangement of the PPN parameters within the standard metric [cf. \eqref{metricPPN00}--\eqref{metricPPNij} with $\beta_1=\beta$ and $\beta_2=0$] is meticulously structured to enable a clear physical interpretation of each parameter, linking them to distinct properties or measurable phenomena such as time delay, bending of light by a massive object, etc. This systematic organization allows for a deeper understanding of the gravitational effects encapsulated by the PPN formalism. For a concise overview, Table \ref{tab:parameters2} presents the significance of each PPN parameter, while the current constraints on these parameters are detailed in Table \ref{tab:bounds3}. Such tables serve as a valuable reference for evaluating the implications of the PPN framework in a gravity theory.
\begin{table}[t!]
  \caption{The PPN Parameters physical meaning.}
  \label{tab:parameters2}
  \renewcommand{\arraystretch}{1.5}
  \centering
  \begin{tabular}{p{1.5 cm}|p{4.2 cm}|c}
    \hline \hline
    Parameter & Physical meaning associated & Value in GR \\
    \hline
    $\gamma$ & light motion tests & 1\\
    \hline
    $\beta$ & Periastron shift in a binary system & 1\\
    \hline
    $\xi$ & Preferred-location effects & 0\\
    \hline
    $\alpha_1$, $\alpha_2$ & Preferred-frame effects & all null\\
    and $\alpha_3$ & \\
    \hline
    $\zeta_1$, $\zeta_2$, $\zeta_3$, & Violation of conservation & all null \\
    $\zeta_4$ and $\alpha_3$ & of total momentum & \\
    \hline \hline
  \end{tabular}
  \renewcommand{\arraystretch}{1.0}
\end{table}

\vspace{3cm}

\begin{table}[t!]
  \caption[Current limits on the PPN parameters.]{Current limits on PPN parameters (from Ref. \cite{Park:2017zgd} for $\beta$ and Ref. \cite{Will:1971zzb} for the others). }
  \label{tab:bounds3}
  \renewcommand{\arraystretch}{1.5}
  \centering
  \begin{tabular}{c|r||c|c}
    \hline \hline
    Parameter & Limit\phantom{AA} & 
    Parameter & Limit \\
    \hline
    $\gamma-1$ & \ ${\scriptstyle \lesssim\,} 2.3 \times 10^{-5 \phantom{0}}$ & $\alpha_3$ &   \ ${\scriptstyle \lesssim\,}4 \times 10^{-20}$ \\
    $\beta-1$ & \ ${\scriptstyle \lesssim\,}2.7 \times 10^{-5 \phantom{0}}$ &$\zeta_1$ &  \ ${\scriptstyle \lesssim\,}2 \times 10^{-2 \phantom{0}}$ \\
    $\xi$ & \ ${\scriptstyle \lesssim\,}4 \times  10^{-9 \phantom{0}}$ & $\zeta_2$ & \ ${\scriptstyle \lesssim\,}4 \times 10^{-5 \phantom{0}}$ \\
    $\alpha_1$ & \ ${\scriptstyle \lesssim\,}7 \times 10^{-5 \phantom{0}}$ & $\zeta_3$ & \ ${\scriptstyle \lesssim\,}10^{-8 \phantom{0}}$ \\
    $\alpha_2$ & \ ${\scriptstyle \lesssim\,}2 \times 10^{-9 \phantom{0}}$ & $\zeta_4$ &  \multicolumn{1}{c}{---} \\
    \hline \hline
  \end{tabular}
\end{table}

\bigskip
\bigskip

\bibliography{AllMyRefs}{} 

\begin{thebibliography}{64}%
\makeatletter
\providecommand \@ifxundefined [1]{%
 \@ifx{#1\undefined}
}%
\providecommand \@ifnum [1]{%
 \ifnum #1\expandafter \@firstoftwo
 \else \expandafter \@secondoftwo
 \fi
}%
\providecommand \@ifx [1]{%
 \ifx #1\expandafter \@firstoftwo
 \else \expandafter \@secondoftwo
 \fi
}%
\providecommand \natexlab [1]{#1}%
\providecommand \enquote  [1]{``#1''}%
\providecommand \bibnamefont  [1]{#1}%
\providecommand \bibfnamefont [1]{#1}%
\providecommand \citenamefont [1]{#1}%
\providecommand \href@noop [0]{\@secondoftwo}%
\providecommand \href [0]{\begingroup \@sanitize@url \@href}%
\providecommand \@href[1]{\@@startlink{#1}\@@href}%
\providecommand \@@href[1]{\endgroup#1\@@endlink}%
\providecommand \@sanitize@url [0]{\catcode `\\12\catcode `\$12\catcode `\&12\catcode `\#12\catcode `\^12\catcode `\_12\catcode `\%12\relax}%
\providecommand \@@startlink[1]{}%
\providecommand \@@endlink[0]{}%
\providecommand \url  [0]{\begingroup\@sanitize@url \@url }%
\providecommand \@url [1]{\endgroup\@href {#1}{\urlprefix }}%
\providecommand \urlprefix  [0]{URL }%
\providecommand \Eprint [0]{\href }%
\providecommand \doibase [0]{https://doi.org/}%
\providecommand \selectlanguage [0]{\@gobble}%
\providecommand \bibinfo  [0]{\@secondoftwo}%
\providecommand \bibfield  [0]{\@secondoftwo}%
\providecommand \translation [1]{[#1]}%
\providecommand \BibitemOpen [0]{}%
\providecommand \bibitemStop [0]{}%
\providecommand \bibitemNoStop [0]{.\EOS\space}%
\providecommand \EOS [0]{\spacefactor3000\relax}%
\providecommand \BibitemShut  [1]{\csname bibitem#1\endcsname}%
\let\auto@bib@innerbib\@empty
\bibitem [{\citenamefont {Damour}\ and\ \citenamefont {Esposito-Farese}(1993)}]{Damour:1993hw}%
  \BibitemOpen
  \bibfield  {author} {\bibinfo {author} {\bibfnamefont {T.}~\bibnamefont {Damour}}\ and\ \bibinfo {author} {\bibfnamefont {G.}~\bibnamefont {Esposito-Farese}},\ }\bibfield  {title} {\bibinfo {title} {{Nonperturbative strong field effects in tensor - scalar theories of gravitation}},\ }\href {https://doi.org/10.1103/PhysRevLett.70.2220} {\bibfield  {journal} {\bibinfo  {journal} {Phys. Rev. Lett.}\ }\textbf {\bibinfo {volume} {70}},\ \bibinfo {pages} {2220} (\bibinfo {year} {1993})}\BibitemShut {NoStop}%
\bibitem [{\citenamefont {Doneva}\ \emph {et~al.}(2024)\citenamefont {Doneva}, \citenamefont {Ramazano\u{g}lu}, \citenamefont {Silva}, \citenamefont {Sotiriou},\ and\ \citenamefont {Yazadjiev}}]{Doneva:2022ewd}%
  \BibitemOpen
  \bibfield  {author} {\bibinfo {author} {\bibfnamefont {D.~D.}\ \bibnamefont {Doneva}}, \bibinfo {author} {\bibfnamefont {F.~M.}\ \bibnamefont {Ramazano\u{g}lu}}, \bibinfo {author} {\bibfnamefont {H.~O.}\ \bibnamefont {Silva}}, \bibinfo {author} {\bibfnamefont {T.~P.}\ \bibnamefont {Sotiriou}},\ and\ \bibinfo {author} {\bibfnamefont {S.~S.}\ \bibnamefont {Yazadjiev}},\ }\bibfield  {title} {\bibinfo {title} {{Spontaneous scalarization}},\ }\href {https://doi.org/10.1103/RevModPhys.96.015004} {\bibfield  {journal} {\bibinfo  {journal} {Rev. Mod. Phys.}\ }\textbf {\bibinfo {volume} {96}},\ \bibinfo {pages} {015004} (\bibinfo {year} {2024})},\ \Eprint {https://arxiv.org/abs/2211.01766} {arXiv:2211.01766 [gr-qc]} \BibitemShut {NoStop}%
\bibitem [{\citenamefont {Antoniou}\ \emph {et~al.}(2018)\citenamefont {Antoniou}, \citenamefont {Bakopoulos},\ and\ \citenamefont {Kanti}}]{Antoniou:2017acq}%
  \BibitemOpen
  \bibfield  {author} {\bibinfo {author} {\bibfnamefont {G.}~\bibnamefont {Antoniou}}, \bibinfo {author} {\bibfnamefont {A.}~\bibnamefont {Bakopoulos}},\ and\ \bibinfo {author} {\bibfnamefont {P.}~\bibnamefont {Kanti}},\ }\bibfield  {title} {\bibinfo {title} {{Evasion of No-Hair Theorems and Novel Black-Hole Solutions in Gauss-Bonnet Theories}},\ }\href {https://doi.org/10.1103/PhysRevLett.120.131102} {\bibfield  {journal} {\bibinfo  {journal} {Phys. Rev. Lett.}\ }\textbf {\bibinfo {volume} {120}},\ \bibinfo {pages} {131102} (\bibinfo {year} {2018})},\ \Eprint {https://arxiv.org/abs/1711.03390} {arXiv:1711.03390 [hep-th]} \BibitemShut {NoStop}%
\bibitem [{\citenamefont {Herdeiro}\ \emph {et~al.}(2018)\citenamefont {Herdeiro}, \citenamefont {Radu}, \citenamefont {Sanchis-Gual},\ and\ \citenamefont {Font}}]{Herdeiro:2018wub}%
  \BibitemOpen
  \bibfield  {author} {\bibinfo {author} {\bibfnamefont {C.~A.~R.}\ \bibnamefont {Herdeiro}}, \bibinfo {author} {\bibfnamefont {E.}~\bibnamefont {Radu}}, \bibinfo {author} {\bibfnamefont {N.}~\bibnamefont {Sanchis-Gual}},\ and\ \bibinfo {author} {\bibfnamefont {J.~A.}\ \bibnamefont {Font}},\ }\bibfield  {title} {\bibinfo {title} {{Spontaneous Scalarization of Charged Black Holes}},\ }\href {https://doi.org/10.1103/PhysRevLett.121.101102} {\bibfield  {journal} {\bibinfo  {journal} {Phys. Rev. Lett.}\ }\textbf {\bibinfo {volume} {121}},\ \bibinfo {pages} {101102} (\bibinfo {year} {2018})},\ \Eprint {https://arxiv.org/abs/1806.05190} {arXiv:1806.05190 [gr-qc]} \BibitemShut {NoStop}%
\bibitem [{\citenamefont {Dima}\ \emph {et~al.}(2020)\citenamefont {Dima}, \citenamefont {Barausse}, \citenamefont {Franchini},\ and\ \citenamefont {Sotiriou}}]{Dima:2020yac}%
  \BibitemOpen
  \bibfield  {author} {\bibinfo {author} {\bibfnamefont {A.}~\bibnamefont {Dima}}, \bibinfo {author} {\bibfnamefont {E.}~\bibnamefont {Barausse}}, \bibinfo {author} {\bibfnamefont {N.}~\bibnamefont {Franchini}},\ and\ \bibinfo {author} {\bibfnamefont {T.~P.}\ \bibnamefont {Sotiriou}},\ }\bibfield  {title} {\bibinfo {title} {{Spin-induced black hole spontaneous scalarization}},\ }\href {https://doi.org/10.1103/PhysRevLett.125.231101} {\bibfield  {journal} {\bibinfo  {journal} {Phys. Rev. Lett.}\ }\textbf {\bibinfo {volume} {125}},\ \bibinfo {pages} {231101} (\bibinfo {year} {2020})},\ \Eprint {https://arxiv.org/abs/2006.03095} {arXiv:2006.03095 [gr-qc]} \BibitemShut {NoStop}%
\bibitem [{\citenamefont {East}\ and\ \citenamefont {Ripley}(2021)}]{East:2021bqk}%
  \BibitemOpen
  \bibfield  {author} {\bibinfo {author} {\bibfnamefont {W.~E.}\ \bibnamefont {East}}\ and\ \bibinfo {author} {\bibfnamefont {J.~L.}\ \bibnamefont {Ripley}},\ }\bibfield  {title} {\bibinfo {title} {{Dynamics of Spontaneous Black Hole Scalarization and Mergers in Einstein-Scalar-Gauss-Bonnet Gravity}},\ }\href {https://doi.org/10.1103/PhysRevLett.127.101102} {\bibfield  {journal} {\bibinfo  {journal} {Phys. Rev. Lett.}\ }\textbf {\bibinfo {volume} {127}},\ \bibinfo {pages} {101102} (\bibinfo {year} {2021})},\ \Eprint {https://arxiv.org/abs/2105.08571} {arXiv:2105.08571 [gr-qc]} \BibitemShut {NoStop}%
\bibitem [{\citenamefont {Berti}\ \emph {et~al.}(2021)\citenamefont {Berti}, \citenamefont {Collodel}, \citenamefont {Kleihaus},\ and\ \citenamefont {Kunz}}]{Berti:2020kgk}%
  \BibitemOpen
  \bibfield  {author} {\bibinfo {author} {\bibfnamefont {E.}~\bibnamefont {Berti}}, \bibinfo {author} {\bibfnamefont {L.~G.}\ \bibnamefont {Collodel}}, \bibinfo {author} {\bibfnamefont {B.}~\bibnamefont {Kleihaus}},\ and\ \bibinfo {author} {\bibfnamefont {J.}~\bibnamefont {Kunz}},\ }\bibfield  {title} {\bibinfo {title} {{Spin-induced black-hole scalarization in Einstein-scalar-Gauss-Bonnet theory}},\ }\href {https://doi.org/10.1103/PhysRevLett.126.011104} {\bibfield  {journal} {\bibinfo  {journal} {Phys. Rev. Lett.}\ }\textbf {\bibinfo {volume} {126}},\ \bibinfo {pages} {011104} (\bibinfo {year} {2021})},\ \Eprint {https://arxiv.org/abs/2009.03905} {arXiv:2009.03905 [gr-qc]} \BibitemShut {NoStop}%
\bibitem [{\citenamefont {Zhang}\ \emph {et~al.}(2022)\citenamefont {Zhang}, \citenamefont {Chen}, \citenamefont {Liu}, \citenamefont {Luo}, \citenamefont {Tian},\ and\ \citenamefont {Wang}}]{Zhang:2021nnn}%
  \BibitemOpen
  \bibfield  {author} {\bibinfo {author} {\bibfnamefont {C.-Y.}\ \bibnamefont {Zhang}}, \bibinfo {author} {\bibfnamefont {Q.}~\bibnamefont {Chen}}, \bibinfo {author} {\bibfnamefont {Y.}~\bibnamefont {Liu}}, \bibinfo {author} {\bibfnamefont {W.-K.}\ \bibnamefont {Luo}}, \bibinfo {author} {\bibfnamefont {Y.}~\bibnamefont {Tian}},\ and\ \bibinfo {author} {\bibfnamefont {B.}~\bibnamefont {Wang}},\ }\bibfield  {title} {\bibinfo {title} {{Critical Phenomena in Dynamical Scalarization of Charged Black Holes}},\ }\href {https://doi.org/10.1103/PhysRevLett.128.161105} {\bibfield  {journal} {\bibinfo  {journal} {Phys. Rev. Lett.}\ }\textbf {\bibinfo {volume} {128}},\ \bibinfo {pages} {161105} (\bibinfo {year} {2022})},\ \Eprint {https://arxiv.org/abs/2112.07455} {arXiv:2112.07455 [gr-qc]} \BibitemShut {NoStop}%
\bibitem [{\citenamefont {Silva}\ \emph {et~al.}(2021)\citenamefont {Silva}, \citenamefont {Witek}, \citenamefont {Elley},\ and\ \citenamefont {Yunes}}]{Silva:2020omi}%
  \BibitemOpen
  \bibfield  {author} {\bibinfo {author} {\bibfnamefont {H.~O.}\ \bibnamefont {Silva}}, \bibinfo {author} {\bibfnamefont {H.}~\bibnamefont {Witek}}, \bibinfo {author} {\bibfnamefont {M.}~\bibnamefont {Elley}},\ and\ \bibinfo {author} {\bibfnamefont {N.}~\bibnamefont {Yunes}},\ }\bibfield  {title} {\bibinfo {title} {{Dynamical Descalarization in Binary Black Hole Mergers}},\ }\href {https://doi.org/10.1103/PhysRevLett.127.031101} {\bibfield  {journal} {\bibinfo  {journal} {Phys. Rev. Lett.}\ }\textbf {\bibinfo {volume} {127}},\ \bibinfo {pages} {031101} (\bibinfo {year} {2021})},\ \Eprint {https://arxiv.org/abs/2012.10436} {arXiv:2012.10436 [gr-qc]} \BibitemShut {NoStop}%
\bibitem [{\citenamefont {Herdeiro}\ \emph {et~al.}(2021)\citenamefont {Herdeiro}, \citenamefont {Radu}, \citenamefont {Silva}, \citenamefont {Sotiriou},\ and\ \citenamefont {Yunes}}]{Herdeiro:2020wei}%
  \BibitemOpen
  \bibfield  {author} {\bibinfo {author} {\bibfnamefont {C.~A.~R.}\ \bibnamefont {Herdeiro}}, \bibinfo {author} {\bibfnamefont {E.}~\bibnamefont {Radu}}, \bibinfo {author} {\bibfnamefont {H.~O.}\ \bibnamefont {Silva}}, \bibinfo {author} {\bibfnamefont {T.~P.}\ \bibnamefont {Sotiriou}},\ and\ \bibinfo {author} {\bibfnamefont {N.}~\bibnamefont {Yunes}},\ }\bibfield  {title} {\bibinfo {title} {{Spin-induced scalarized black holes}},\ }\href {https://doi.org/10.1103/PhysRevLett.126.011103} {\bibfield  {journal} {\bibinfo  {journal} {Phys. Rev. Lett.}\ }\textbf {\bibinfo {volume} {126}},\ \bibinfo {pages} {011103} (\bibinfo {year} {2021})},\ \Eprint {https://arxiv.org/abs/2009.03904} {arXiv:2009.03904 [gr-qc]} \BibitemShut {NoStop}%
\bibitem [{\citenamefont {Cunha}\ \emph {et~al.}(2019)\citenamefont {Cunha}, \citenamefont {Herdeiro},\ and\ \citenamefont {Radu}}]{Cunha:2019dwb}%
  \BibitemOpen
  \bibfield  {author} {\bibinfo {author} {\bibfnamefont {P.~V.~P.}\ \bibnamefont {Cunha}}, \bibinfo {author} {\bibfnamefont {C.~A.~R.}\ \bibnamefont {Herdeiro}},\ and\ \bibinfo {author} {\bibfnamefont {E.}~\bibnamefont {Radu}},\ }\bibfield  {title} {\bibinfo {title} {{Spontaneously Scalarized Kerr Black Holes in Extended Scalar-Tensor\textendash{}Gauss-Bonnet Gravity}},\ }\href {https://doi.org/10.1103/PhysRevLett.123.011101} {\bibfield  {journal} {\bibinfo  {journal} {Phys. Rev. Lett.}\ }\textbf {\bibinfo {volume} {123}},\ \bibinfo {pages} {011101} (\bibinfo {year} {2019})},\ \Eprint {https://arxiv.org/abs/1904.09997} {arXiv:1904.09997 [gr-qc]} \BibitemShut {NoStop}%
\bibitem [{\citenamefont {Silva}\ \emph {et~al.}(2018)\citenamefont {Silva}, \citenamefont {Sakstein}, \citenamefont {Gualtieri}, \citenamefont {Sotiriou},\ and\ \citenamefont {Berti}}]{Silva:2017uqg}%
  \BibitemOpen
  \bibfield  {author} {\bibinfo {author} {\bibfnamefont {H.~O.}\ \bibnamefont {Silva}}, \bibinfo {author} {\bibfnamefont {J.}~\bibnamefont {Sakstein}}, \bibinfo {author} {\bibfnamefont {L.}~\bibnamefont {Gualtieri}}, \bibinfo {author} {\bibfnamefont {T.~P.}\ \bibnamefont {Sotiriou}},\ and\ \bibinfo {author} {\bibfnamefont {E.}~\bibnamefont {Berti}},\ }\bibfield  {title} {\bibinfo {title} {{Spontaneous scalarization of black holes and compact stars from a Gauss-Bonnet coupling}},\ }\href {https://doi.org/10.1103/PhysRevLett.120.131104} {\bibfield  {journal} {\bibinfo  {journal} {Phys. Rev. Lett.}\ }\textbf {\bibinfo {volume} {120}},\ \bibinfo {pages} {131104} (\bibinfo {year} {2018})},\ \Eprint {https://arxiv.org/abs/1711.02080} {arXiv:1711.02080 [gr-qc]} \BibitemShut {NoStop}%
\bibitem [{\citenamefont {Mendes}\ and\ \citenamefont {Ortiz}(2018)}]{Mendes:2018qwo}%
  \BibitemOpen
  \bibfield  {author} {\bibinfo {author} {\bibfnamefont {R.~F.~P.}\ \bibnamefont {Mendes}}\ and\ \bibinfo {author} {\bibfnamefont {N.}~\bibnamefont {Ortiz}},\ }\bibfield  {title} {\bibinfo {title} {{New class of quasinormal modes of neutron stars in scalar-tensor gravity}},\ }\href {https://doi.org/10.1103/PhysRevLett.120.201104} {\bibfield  {journal} {\bibinfo  {journal} {Phys. Rev. Lett.}\ }\textbf {\bibinfo {volume} {120}},\ \bibinfo {pages} {201104} (\bibinfo {year} {2018})},\ \Eprint {https://arxiv.org/abs/1802.07847} {arXiv:1802.07847 [gr-qc]} \BibitemShut {NoStop}%
\bibitem [{\citenamefont {Kuan}\ \emph {et~al.}(2021)\citenamefont {Kuan}, \citenamefont {Doneva},\ and\ \citenamefont {Yazadjiev}}]{Kuan:2021lol}%
  \BibitemOpen
  \bibfield  {author} {\bibinfo {author} {\bibfnamefont {H.-J.}\ \bibnamefont {Kuan}}, \bibinfo {author} {\bibfnamefont {D.~D.}\ \bibnamefont {Doneva}},\ and\ \bibinfo {author} {\bibfnamefont {S.~S.}\ \bibnamefont {Yazadjiev}},\ }\bibfield  {title} {\bibinfo {title} {{Dynamical Formation of Scalarized Black Holes and Neutron Stars through Stellar Core Collapse}},\ }\href {https://doi.org/10.1103/PhysRevLett.127.161103} {\bibfield  {journal} {\bibinfo  {journal} {Phys. Rev. Lett.}\ }\textbf {\bibinfo {volume} {127}},\ \bibinfo {pages} {161103} (\bibinfo {year} {2021})},\ \Eprint {https://arxiv.org/abs/2103.11999} {arXiv:2103.11999 [gr-qc]} \BibitemShut {NoStop}%
\bibitem [{\citenamefont {Clifton}\ \emph {et~al.}(2012)\citenamefont {Clifton}, \citenamefont {Ferreira}, \citenamefont {Padilla},\ and\ \citenamefont {Skordis}}]{Clifton:2011jh}%
  \BibitemOpen
  \bibfield  {author} {\bibinfo {author} {\bibfnamefont {T.}~\bibnamefont {Clifton}}, \bibinfo {author} {\bibfnamefont {P.~G.}\ \bibnamefont {Ferreira}}, \bibinfo {author} {\bibfnamefont {A.}~\bibnamefont {Padilla}},\ and\ \bibinfo {author} {\bibfnamefont {C.}~\bibnamefont {Skordis}},\ }\bibfield  {title} {\bibinfo {title} {{Modified Gravity and Cosmology}},\ }\href {https://doi.org/10.1016/j.physrep.2012.01.001} {\bibfield  {journal} {\bibinfo  {journal} {Phys. Rept.}\ }\textbf {\bibinfo {volume} {513}},\ \bibinfo {pages} {1} (\bibinfo {year} {2012})},\ \Eprint {https://arxiv.org/abs/1106.2476} {arXiv:1106.2476 [astro-ph.CO]} \BibitemShut {NoStop}%
\bibitem [{\citenamefont {Koyama}(2016)}]{Koyama:2015vza}%
  \BibitemOpen
  \bibfield  {author} {\bibinfo {author} {\bibfnamefont {K.}~\bibnamefont {Koyama}},\ }\bibfield  {title} {\bibinfo {title} {{Cosmological Tests of Modified Gravity}},\ }\href {https://doi.org/10.1088/0034-4885/79/4/046902} {\bibfield  {journal} {\bibinfo  {journal} {Rept. Prog. Phys.}\ }\textbf {\bibinfo {volume} {79}},\ \bibinfo {pages} {046902} (\bibinfo {year} {2016})},\ \Eprint {https://arxiv.org/abs/1504.04623} {arXiv:1504.04623 [astro-ph.CO]} \BibitemShut {NoStop}%
\bibitem [{\citenamefont {Joyce}\ \emph {et~al.}(2015)\citenamefont {Joyce}, \citenamefont {Jain}, \citenamefont {Khoury},\ and\ \citenamefont {Trodden}}]{Joyce:2014kja}%
  \BibitemOpen
  \bibfield  {author} {\bibinfo {author} {\bibfnamefont {A.}~\bibnamefont {Joyce}}, \bibinfo {author} {\bibfnamefont {B.}~\bibnamefont {Jain}}, \bibinfo {author} {\bibfnamefont {J.}~\bibnamefont {Khoury}},\ and\ \bibinfo {author} {\bibfnamefont {M.}~\bibnamefont {Trodden}},\ }\bibfield  {title} {\bibinfo {title} {{Beyond the Cosmological Standard Model}},\ }\href {https://doi.org/10.1016/j.physrep.2014.12.002} {\bibfield  {journal} {\bibinfo  {journal} {Phys. Rept.}\ }\textbf {\bibinfo {volume} {568}},\ \bibinfo {pages} {1} (\bibinfo {year} {2015})},\ \Eprint {https://arxiv.org/abs/1407.0059} {arXiv:1407.0059 [astro-ph.CO]} \BibitemShut {NoStop}%
\bibitem [{\citenamefont {Ishak}(2019)}]{Ishak:2018his}%
  \BibitemOpen
  \bibfield  {author} {\bibinfo {author} {\bibfnamefont {M.}~\bibnamefont {Ishak}},\ }\bibfield  {title} {\bibinfo {title} {{Testing General Relativity in Cosmology}},\ }\href {https://doi.org/10.1007/s41114-018-0017-4} {\bibfield  {journal} {\bibinfo  {journal} {Living Rev. Rel.}\ }\textbf {\bibinfo {volume} {22}},\ \bibinfo {pages} {1} (\bibinfo {year} {2019})},\ \Eprint {https://arxiv.org/abs/1806.10122} {arXiv:1806.10122 [astro-ph.CO]} \BibitemShut {NoStop}%
\bibitem [{\citenamefont {Doneva}\ and\ \citenamefont {Yazadjiev}(2018)}]{Doneva:2017bvd}%
  \BibitemOpen
  \bibfield  {author} {\bibinfo {author} {\bibfnamefont {D.~D.}\ \bibnamefont {Doneva}}\ and\ \bibinfo {author} {\bibfnamefont {S.~S.}\ \bibnamefont {Yazadjiev}},\ }\bibfield  {title} {\bibinfo {title} {{New Gauss-Bonnet Black Holes with Curvature-Induced Scalarization in Extended Scalar-Tensor Theories}},\ }\href {https://doi.org/10.1103/PhysRevLett.120.131103} {\bibfield  {journal} {\bibinfo  {journal} {Phys. Rev. Lett.}\ }\textbf {\bibinfo {volume} {120}},\ \bibinfo {pages} {131103} (\bibinfo {year} {2018})},\ \Eprint {https://arxiv.org/abs/1711.01187} {arXiv:1711.01187 [gr-qc]} \BibitemShut {NoStop}%
\bibitem [{\citenamefont {Ramazano\u{g}lu}(2017)}]{Ramazanoglu:2017xbl}%
  \BibitemOpen
  \bibfield  {author} {\bibinfo {author} {\bibfnamefont {F.~M.}\ \bibnamefont {Ramazano\u{g}lu}},\ }\bibfield  {title} {\bibinfo {title} {{Spontaneous growth of vector fields in gravity}},\ }\href {https://doi.org/10.1103/PhysRevD.96.064009} {\bibfield  {journal} {\bibinfo  {journal} {Phys. Rev. D}\ }\textbf {\bibinfo {volume} {96}},\ \bibinfo {pages} {064009} (\bibinfo {year} {2017})},\ \Eprint {https://arxiv.org/abs/1706.01056} {arXiv:1706.01056 [gr-qc]} \BibitemShut {NoStop}%
\bibitem [{\citenamefont {Annulli}\ \emph {et~al.}(2019)\citenamefont {Annulli}, \citenamefont {Cardoso},\ and\ \citenamefont {Gualtieri}}]{Annulli:2019fzq}%
  \BibitemOpen
  \bibfield  {author} {\bibinfo {author} {\bibfnamefont {L.}~\bibnamefont {Annulli}}, \bibinfo {author} {\bibfnamefont {V.}~\bibnamefont {Cardoso}},\ and\ \bibinfo {author} {\bibfnamefont {L.}~\bibnamefont {Gualtieri}},\ }\bibfield  {title} {\bibinfo {title} {{Electromagnetism and hidden vector fields in modified gravity theories: spontaneous and induced vectorization}},\ }\href {https://doi.org/10.1103/PhysRevD.99.044038} {\bibfield  {journal} {\bibinfo  {journal} {Phys. Rev. D}\ }\textbf {\bibinfo {volume} {99}},\ \bibinfo {pages} {044038} (\bibinfo {year} {2019})},\ \Eprint {https://arxiv.org/abs/1901.02461} {arXiv:1901.02461 [gr-qc]} \BibitemShut {NoStop}%
\bibitem [{\citenamefont {Minamitsuji}(2020)}]{Minamitsuji:2020pak}%
  \BibitemOpen
  \bibfield  {author} {\bibinfo {author} {\bibfnamefont {M.}~\bibnamefont {Minamitsuji}},\ }\bibfield  {title} {\bibinfo {title} {{Spontaneous vectorization in the presence of vector field coupling to matter}},\ }\href {https://doi.org/10.1103/PhysRevD.101.104044} {\bibfield  {journal} {\bibinfo  {journal} {Phys. Rev. D}\ }\textbf {\bibinfo {volume} {101}},\ \bibinfo {pages} {104044} (\bibinfo {year} {2020})},\ \Eprint {https://arxiv.org/abs/2003.11885} {arXiv:2003.11885 [gr-qc]} \BibitemShut {NoStop}%
\bibitem [{\citenamefont {Kase}\ \emph {et~al.}(2020)\citenamefont {Kase}, \citenamefont {Minamitsuji},\ and\ \citenamefont {Tsujikawa}}]{Kase:2020yhw}%
  \BibitemOpen
  \bibfield  {author} {\bibinfo {author} {\bibfnamefont {R.}~\bibnamefont {Kase}}, \bibinfo {author} {\bibfnamefont {M.}~\bibnamefont {Minamitsuji}},\ and\ \bibinfo {author} {\bibfnamefont {S.}~\bibnamefont {Tsujikawa}},\ }\bibfield  {title} {\bibinfo {title} {{Neutron stars with a generalized Proca hair and spontaneous vectorization}},\ }\href {https://doi.org/10.1103/PhysRevD.102.024067} {\bibfield  {journal} {\bibinfo  {journal} {Phys. Rev. D}\ }\textbf {\bibinfo {volume} {102}},\ \bibinfo {pages} {024067} (\bibinfo {year} {2020})},\ \Eprint {https://arxiv.org/abs/2001.10701} {arXiv:2001.10701 [gr-qc]} \BibitemShut {NoStop}%
\bibitem [{\citenamefont {Ye}\ \emph {et~al.}(2024)\citenamefont {Ye}, \citenamefont {Chen}, \citenamefont {Fu}, \citenamefont {Niu}, \citenamefont {Zhang},\ and\ \citenamefont {Liu}}]{Ye:2024pyy}%
  \BibitemOpen
  \bibfield  {author} {\bibinfo {author} {\bibfnamefont {G.-Z.}\ \bibnamefont {Ye}}, \bibinfo {author} {\bibfnamefont {C.-Y.}\ \bibnamefont {Chen}}, \bibinfo {author} {\bibfnamefont {G.}~\bibnamefont {Fu}}, \bibinfo {author} {\bibfnamefont {C.}~\bibnamefont {Niu}}, \bibinfo {author} {\bibfnamefont {C.-Y.}\ \bibnamefont {Zhang}},\ and\ \bibinfo {author} {\bibfnamefont {P.}~\bibnamefont {Liu}},\ }\bibfield  {title} {\bibinfo {title} {{Spontaneous Vectorization in the Einstein-Maxwell-Vector Model}},\ }\Eprint {https://arxiv.org/abs/2410.16920} {arXiv:2410.16920 [gr-qc]}  (\bibinfo {year} {2024})\BibitemShut {NoStop}%
\bibitem [{\citenamefont {Oliveira}\ and\ \citenamefont {Pombo}(2021)}]{Oliveira:2020dru}%
  \BibitemOpen
  \bibfield  {author} {\bibinfo {author} {\bibfnamefont {J.~a. M.~S.}\ \bibnamefont {Oliveira}}\ and\ \bibinfo {author} {\bibfnamefont {A.~M.}\ \bibnamefont {Pombo}},\ }\bibfield  {title} {\bibinfo {title} {{Spontaneous vectorization of electrically charged black holes}},\ }\href {https://doi.org/10.1103/PhysRevD.103.044004} {\bibfield  {journal} {\bibinfo  {journal} {Phys. Rev. D}\ }\textbf {\bibinfo {volume} {103}},\ \bibinfo {pages} {044004} (\bibinfo {year} {2021})},\ \Eprint {https://arxiv.org/abs/2012.07869} {arXiv:2012.07869 [gr-qc]} \BibitemShut {NoStop}%
\bibitem [{\citenamefont {Ramazano\u{g}lu}(2019)}]{Ramazanoglu:2019gbz}%
  \BibitemOpen
  \bibfield  {author} {\bibinfo {author} {\bibfnamefont {F.~M.}\ \bibnamefont {Ramazano\u{g}lu}},\ }\bibfield  {title} {\bibinfo {title} {{Spontaneous tensorization from curvature coupling and beyond}},\ }\href {https://doi.org/10.1103/PhysRevD.99.084015} {\bibfield  {journal} {\bibinfo  {journal} {Phys. Rev. D}\ }\textbf {\bibinfo {volume} {99}},\ \bibinfo {pages} {084015} (\bibinfo {year} {2019})},\ \Eprint {https://arxiv.org/abs/1901.10009} {arXiv:1901.10009 [gr-qc]} \BibitemShut {NoStop}%
\bibitem [{\citenamefont {Ramazano\u{g}lu}\ and\ \citenamefont {\"Unl\"ut\"urk}(2019)}]{Ramazanoglu:2019jrr}%
  \BibitemOpen
  \bibfield  {author} {\bibinfo {author} {\bibfnamefont {F.~M.}\ \bibnamefont {Ramazano\u{g}lu}}\ and\ \bibinfo {author} {\bibfnamefont {K.~I.}\ \bibnamefont {\"Unl\"ut\"urk}},\ }\bibfield  {title} {\bibinfo {title} {{Generalized disformal coupling leads to spontaneous tensorization}},\ }\href {https://doi.org/10.1103/PhysRevD.100.084026} {\bibfield  {journal} {\bibinfo  {journal} {Phys. Rev. D}\ }\textbf {\bibinfo {volume} {100}},\ \bibinfo {pages} {084026} (\bibinfo {year} {2019})},\ \Eprint {https://arxiv.org/abs/1910.02801} {arXiv:1910.02801 [gr-qc]} \BibitemShut {NoStop}%
\bibitem [{\citenamefont {Silva}\ \emph {et~al.}(2022)\citenamefont {Silva}, \citenamefont {Coates}, \citenamefont {Ramazano\u{g}lu},\ and\ \citenamefont {Sotiriou}}]{Silva:2021jya}%
  \BibitemOpen
  \bibfield  {author} {\bibinfo {author} {\bibfnamefont {H.~O.}\ \bibnamefont {Silva}}, \bibinfo {author} {\bibfnamefont {A.}~\bibnamefont {Coates}}, \bibinfo {author} {\bibfnamefont {F.~M.}\ \bibnamefont {Ramazano\u{g}lu}},\ and\ \bibinfo {author} {\bibfnamefont {T.~P.}\ \bibnamefont {Sotiriou}},\ }\bibfield  {title} {\bibinfo {title} {{Ghost of vector fields in compact stars}},\ }\href {https://doi.org/10.1103/PhysRevD.105.024046} {\bibfield  {journal} {\bibinfo  {journal} {Phys. Rev. D}\ }\textbf {\bibinfo {volume} {105}},\ \bibinfo {pages} {024046} (\bibinfo {year} {2022})},\ \Eprint {https://arxiv.org/abs/2110.04594} {arXiv:2110.04594 [gr-qc]} \BibitemShut {NoStop}%
\bibitem [{\citenamefont {Wang}\ \emph {et~al.}(2021)\citenamefont {Wang}, \citenamefont {Wu},\ and\ \citenamefont {Yang}}]{Wang:2020ohb}%
  \BibitemOpen
  \bibfield  {author} {\bibinfo {author} {\bibfnamefont {P.}~\bibnamefont {Wang}}, \bibinfo {author} {\bibfnamefont {H.}~\bibnamefont {Wu}},\ and\ \bibinfo {author} {\bibfnamefont {H.}~\bibnamefont {Yang}},\ }\bibfield  {title} {\bibinfo {title} {{Scalarized Einstein-Born-Infeld black holes}},\ }\href {https://doi.org/10.1103/PhysRevD.103.104012} {\bibfield  {journal} {\bibinfo  {journal} {Phys. Rev. D}\ }\textbf {\bibinfo {volume} {103}},\ \bibinfo {pages} {104012} (\bibinfo {year} {2021})},\ \Eprint {https://arxiv.org/abs/2012.01066} {arXiv:2012.01066 [gr-qc]} \BibitemShut {NoStop}%
\bibitem [{\citenamefont {Myung}\ and\ \citenamefont {Zou}(2021)}]{Myung:2020ctt}%
  \BibitemOpen
  \bibfield  {author} {\bibinfo {author} {\bibfnamefont {Y.~S.}\ \bibnamefont {Myung}}\ and\ \bibinfo {author} {\bibfnamefont {D.-C.}\ \bibnamefont {Zou}},\ }\bibfield  {title} {\bibinfo {title} {{Scalarized black holes in the Einstein-Maxwell-scalar theory with a quasitopological term}},\ }\href {https://doi.org/10.1103/PhysRevD.103.024010} {\bibfield  {journal} {\bibinfo  {journal} {Phys. Rev. D}\ }\textbf {\bibinfo {volume} {103}},\ \bibinfo {pages} {024010} (\bibinfo {year} {2021})},\ \Eprint {https://arxiv.org/abs/2011.09665} {arXiv:2011.09665 [gr-qc]} \BibitemShut {NoStop}%
\bibitem [{\citenamefont {Chatzifotis}\ \emph {et~al.}(2022)\citenamefont {Chatzifotis}, \citenamefont {Dorlis}, \citenamefont {Mavromatos},\ and\ \citenamefont {Papantonopoulos}}]{Chatzifotis:2022mob}%
  \BibitemOpen
  \bibfield  {author} {\bibinfo {author} {\bibfnamefont {N.}~\bibnamefont {Chatzifotis}}, \bibinfo {author} {\bibfnamefont {P.}~\bibnamefont {Dorlis}}, \bibinfo {author} {\bibfnamefont {N.~E.}\ \bibnamefont {Mavromatos}},\ and\ \bibinfo {author} {\bibfnamefont {E.}~\bibnamefont {Papantonopoulos}},\ }\bibfield  {title} {\bibinfo {title} {{Scalarization of Chern-Simons-Kerr black hole solutions and wormholes}},\ }\href {https://doi.org/10.1103/PhysRevD.105.084051} {\bibfield  {journal} {\bibinfo  {journal} {Phys. Rev. D}\ }\textbf {\bibinfo {volume} {105}},\ \bibinfo {pages} {084051} (\bibinfo {year} {2022})},\ \Eprint {https://arxiv.org/abs/2202.03496} {arXiv:2202.03496 [gr-qc]} \BibitemShut {NoStop}%
\bibitem [{\citenamefont {Antoniou}\ \emph {et~al.}(2021)\citenamefont {Antoniou}, \citenamefont {Bordin},\ and\ \citenamefont {Sotiriou}}]{Antoniou:2020nax}%
  \BibitemOpen
  \bibfield  {author} {\bibinfo {author} {\bibfnamefont {G.}~\bibnamefont {Antoniou}}, \bibinfo {author} {\bibfnamefont {L.}~\bibnamefont {Bordin}},\ and\ \bibinfo {author} {\bibfnamefont {T.~P.}\ \bibnamefont {Sotiriou}},\ }\bibfield  {title} {\bibinfo {title} {{Compact object scalarization with general relativity as a cosmic attractor}},\ }\href {https://doi.org/10.1103/PhysRevD.103.024012} {\bibfield  {journal} {\bibinfo  {journal} {Phys. Rev. D}\ }\textbf {\bibinfo {volume} {103}},\ \bibinfo {pages} {024012} (\bibinfo {year} {2021})},\ \Eprint {https://arxiv.org/abs/2004.14985} {arXiv:2004.14985 [gr-qc]} \BibitemShut {NoStop}%
\bibitem [{\citenamefont {Andreou}\ \emph {et~al.}(2019)\citenamefont {Andreou}, \citenamefont {Franchini}, \citenamefont {Ventagli},\ and\ \citenamefont {Sotiriou}}]{Andreou:2019ikc}%
  \BibitemOpen
  \bibfield  {author} {\bibinfo {author} {\bibfnamefont {N.}~\bibnamefont {Andreou}}, \bibinfo {author} {\bibfnamefont {N.}~\bibnamefont {Franchini}}, \bibinfo {author} {\bibfnamefont {G.}~\bibnamefont {Ventagli}},\ and\ \bibinfo {author} {\bibfnamefont {T.~P.}\ \bibnamefont {Sotiriou}},\ }\bibfield  {title} {\bibinfo {title} {{Spontaneous scalarization in generalised scalar-tensor theory}},\ }\href {https://doi.org/10.1103/PhysRevD.99.124022} {\bibfield  {journal} {\bibinfo  {journal} {Phys. Rev. D}\ }\textbf {\bibinfo {volume} {99}},\ \bibinfo {pages} {124022} (\bibinfo {year} {2019})},\ \bibinfo {note} {[Erratum: Phys.Rev.D 101, 109903 (2020)]},\ \Eprint {https://arxiv.org/abs/1904.06365} {arXiv:1904.06365 [gr-qc]} \BibitemShut {NoStop}%
\bibitem [{\citenamefont {Babichev}\ \emph {et~al.}(2017)\citenamefont {Babichev}, \citenamefont {Charmousis},\ and\ \citenamefont {Leh\'ebel}}]{Babichev:2017guv}%
  \BibitemOpen
  \bibfield  {author} {\bibinfo {author} {\bibfnamefont {E.}~\bibnamefont {Babichev}}, \bibinfo {author} {\bibfnamefont {C.}~\bibnamefont {Charmousis}},\ and\ \bibinfo {author} {\bibfnamefont {A.}~\bibnamefont {Leh\'ebel}},\ }\bibfield  {title} {\bibinfo {title} {{Asymptotically flat black holes in Horndeski theory and beyond}},\ }\href {https://doi.org/10.1088/1475-7516/2017/04/027} {\bibfield  {journal} {\bibinfo  {journal} {JCAP}\ }\textbf {\bibinfo {volume} {04}},\ \bibinfo {pages} {027}},\ \Eprint {https://arxiv.org/abs/1702.01938} {arXiv:1702.01938 [gr-qc]} \BibitemShut {NoStop}%
\bibitem [{\citenamefont {Toniato}\ and\ \citenamefont {Richarte}(2024)}]{Toniato:2024gtx}%
  \BibitemOpen
  \bibfield  {author} {\bibinfo {author} {\bibfnamefont {J.~D.}\ \bibnamefont {Toniato}}\ and\ \bibinfo {author} {\bibfnamefont {M.~G.}\ \bibnamefont {Richarte}},\ }\bibfield  {title} {\bibinfo {title} {{Post-Newtonian analysis of regularized 4D Einstein-Gauss-Bonnet theory: Complete set of PPN parameters and observational constraints}},\ }\href {https://doi.org/10.1103/PhysRevD.109.104068} {\bibfield  {journal} {\bibinfo  {journal} {Phys. Rev. D}\ }\textbf {\bibinfo {volume} {109}},\ \bibinfo {pages} {104068} (\bibinfo {year} {2024})},\ \Eprint {https://arxiv.org/abs/2402.13951} {arXiv:2402.13951 [gr-qc]} \BibitemShut {NoStop}%
\bibitem [{\citenamefont {Will}(1993)}]{Will:1993ns}%
  \BibitemOpen
  \bibfield  {author} {\bibinfo {author} {\bibfnamefont {C.~M.}\ \bibnamefont {Will}},\ }\href@noop {} {\emph {\bibinfo {title} {{Theory and experiment in gravitational physics}}}},\ \bibinfo {edition} {2nd}\ ed.\ (\bibinfo  {publisher} {Cambridge University Press},\ \bibinfo {address} {Cambridge},\ \bibinfo {year} {1993})\BibitemShut {NoStop}%
\bibitem [{\citenamefont {Poisson}\ and\ \citenamefont {Will}(2014)}]{PoissonWill}%
  \BibitemOpen
  \bibfield  {author} {\bibinfo {author} {\bibfnamefont {E.}~\bibnamefont {Poisson}}\ and\ \bibinfo {author} {\bibfnamefont {C.~M.}\ \bibnamefont {Will}},\ }\href@noop {} {\emph {\bibinfo {title} {{Gravity: Newtonian, Post-Newtonian, Relativistic}}}}\ (\bibinfo  {publisher} {Cambridge University Press},\ \bibinfo {year} {2014})\BibitemShut {NoStop}%
\bibitem [{\citenamefont {Will}(1971{\natexlab{a}})}]{Will:1971zza}%
  \BibitemOpen
  \bibfield  {author} {\bibinfo {author} {\bibfnamefont {C.~M.}\ \bibnamefont {Will}},\ }\emph {\bibinfo {title} {{Theoretical frameworks for testing relativistic gravity: The parametrized post-Newtonian formalism}}},\ \href@noop {} {Ph.D. thesis},\ \bibinfo  {school} {Caltech} (\bibinfo {year} {1971}{\natexlab{a}})\BibitemShut {NoStop}%
\bibitem [{\citenamefont {Hohmann}(2021)}]{Hohmann:2020muq}%
  \BibitemOpen
  \bibfield  {author} {\bibinfo {author} {\bibfnamefont {M.}~\bibnamefont {Hohmann}},\ }\bibfield  {title} {\bibinfo {title} {{xPPN: an implementation of the parametrized post-Newtonian formalism using xAct for Mathematica}},\ }\href {https://doi.org/10.1140/epjc/s10052-021-09183-9} {\bibfield  {journal} {\bibinfo  {journal} {Eur. Phys. J. C}\ }\textbf {\bibinfo {volume} {81}},\ \bibinfo {pages} {504} (\bibinfo {year} {2021})},\ \Eprint {https://arxiv.org/abs/2012.14984} {arXiv:2012.14984 [gr-qc]} \BibitemShut {NoStop}%
\bibitem [{\citenamefont {Toniato}\ \emph {et~al.}(2020)\citenamefont {Toniato}, \citenamefont {Rodrigues},\ and\ \citenamefont {Wojnar}}]{Toniato:2019rrd}%
  \BibitemOpen
  \bibfield  {author} {\bibinfo {author} {\bibfnamefont {J.~D.}\ \bibnamefont {Toniato}}, \bibinfo {author} {\bibfnamefont {D.~C.}\ \bibnamefont {Rodrigues}},\ and\ \bibinfo {author} {\bibfnamefont {A.}~\bibnamefont {Wojnar}},\ }\bibfield  {title} {\bibinfo {title} {{Palatini $f(R)$ gravity in the solar system: post-Newtonian equations of motion and complete PPN parameters}},\ }\href {https://doi.org/10.1103/PhysRevD.101.064050} {\bibfield  {journal} {\bibinfo  {journal} {Phys. Rev. D}\ }\textbf {\bibinfo {volume} {101}},\ \bibinfo {pages} {064050} (\bibinfo {year} {2020})},\ \Eprint {https://arxiv.org/abs/1912.12234} {arXiv:1912.12234 [gr-qc]} \BibitemShut {NoStop}%
\bibitem [{\citenamefont {Toniato}\ and\ \citenamefont {Rodrigues}(2021)}]{Toniato:2021vmt}%
  \BibitemOpen
  \bibfield  {author} {\bibinfo {author} {\bibfnamefont {J.~D.}\ \bibnamefont {Toniato}}\ and\ \bibinfo {author} {\bibfnamefont {D.~C.}\ \bibnamefont {Rodrigues}},\ }\bibfield  {title} {\bibinfo {title} {{Post-Newtonian \ensuremath{\gamma}-like parameters and the gravitational slip in scalar-tensor and f(R) theories}},\ }\href {https://doi.org/10.1103/PhysRevD.104.044020} {\bibfield  {journal} {\bibinfo  {journal} {Phys. Rev. D}\ }\textbf {\bibinfo {volume} {104}},\ \bibinfo {pages} {044020} (\bibinfo {year} {2021})},\ \Eprint {https://arxiv.org/abs/2106.12542} {arXiv:2106.12542 [gr-qc]} \BibitemShut {NoStop}%
\bibitem [{\citenamefont {Alves}\ \emph {et~al.}(2024)\citenamefont {Alves}, \citenamefont {Toniato},\ and\ \citenamefont {Rodrigues}}]{alves2024}%
  \BibitemOpen
  \bibfield  {author} {\bibinfo {author} {\bibfnamefont {M.~F.~S.}\ \bibnamefont {Alves}}, \bibinfo {author} {\bibfnamefont {J.~D.}\ \bibnamefont {Toniato}},\ and\ \bibinfo {author} {\bibfnamefont {D.~C.}\ \bibnamefont {Rodrigues}},\ }\bibfield  {title} {\bibinfo {title} {A detailed first-order post-newtonian analysis of massive brans-dicke theories: numerical constraints and the $\beta$ parameter meaning},\ }\href {https://doi.org/10.1103/physrevd.109.044045} {\bibfield  {journal} {\bibinfo  {journal} {Physical Review D}\ }\textbf {\bibinfo {volume} {109}},\ \bibinfo {pages} {044045} (\bibinfo {year} {2024})},\ \Eprint {https://arxiv.org/abs/2307.11883} {arXiv:2307.11883 [gr-qc]} \BibitemShut {NoStop}%
\bibitem [{\citenamefont {Bertotti}\ \emph {et~al.}(2003)\citenamefont {Bertotti}, \citenamefont {Iess},\ and\ \citenamefont {Tortora}}]{Bertotti:2003rm}%
  \BibitemOpen
  \bibfield  {author} {\bibinfo {author} {\bibfnamefont {B.}~\bibnamefont {Bertotti}}, \bibinfo {author} {\bibfnamefont {L.}~\bibnamefont {Iess}},\ and\ \bibinfo {author} {\bibfnamefont {P.}~\bibnamefont {Tortora}},\ }\bibfield  {title} {\bibinfo {title} {{A test of general relativity using radio links with the Cassini spacecraft}},\ }\href {https://doi.org/10.1038/nature01997} {\bibfield  {journal} {\bibinfo  {journal} {Nature}\ }\textbf {\bibinfo {volume} {425}},\ \bibinfo {pages} {374} (\bibinfo {year} {2003})}\BibitemShut {NoStop}%
\bibitem [{\citenamefont {Park}\ \emph {et~al.}(2017)\citenamefont {Park}, \citenamefont {Folkner}, \citenamefont {Konopliv}, \citenamefont {Williams}, \citenamefont {Smith},\ and\ \citenamefont {Zuber}}]{Park:2017zgd}%
  \BibitemOpen
  \bibfield  {author} {\bibinfo {author} {\bibfnamefont {R.~S.}\ \bibnamefont {Park}}, \bibinfo {author} {\bibfnamefont {W.~M.}\ \bibnamefont {Folkner}}, \bibinfo {author} {\bibfnamefont {A.~S.}\ \bibnamefont {Konopliv}}, \bibinfo {author} {\bibfnamefont {J.~G.}\ \bibnamefont {Williams}}, \bibinfo {author} {\bibfnamefont {D.~E.}\ \bibnamefont {Smith}},\ and\ \bibinfo {author} {\bibfnamefont {M.~T.}\ \bibnamefont {Zuber}},\ }\bibfield  {title} {\bibinfo {title} {{Precession of Mercury\textquoteright{}s Perihelion from Ranging to the MESSENGER Spacecraft}},\ }\href {https://doi.org/10.3847/1538-3881/aa5be2} {\bibfield  {journal} {\bibinfo  {journal} {Astron. J.}\ }\textbf {\bibinfo {volume} {153}},\ \bibinfo {pages} {121} (\bibinfo {year} {2017})}\BibitemShut {NoStop}%
\bibitem [{\citenamefont {Hofmann}\ and\ \citenamefont {M\"uller}(2018)}]{Hofmann:2018myc}%
  \BibitemOpen
  \bibfield  {author} {\bibinfo {author} {\bibfnamefont {F.}~\bibnamefont {Hofmann}}\ and\ \bibinfo {author} {\bibfnamefont {J.}~\bibnamefont {M\"uller}},\ }\bibfield  {title} {\bibinfo {title} {{Relativistic tests with lunar laser ranging}},\ }\href {https://doi.org/10.1088/1361-6382/aa8f7a} {\bibfield  {journal} {\bibinfo  {journal} {Class. Quant. Grav.}\ }\textbf {\bibinfo {volume} {35}},\ \bibinfo {pages} {035015} (\bibinfo {year} {2018})}\BibitemShut {NoStop}%
\bibitem [{\citenamefont {Mariani}\ \emph {et~al.}(2024)\citenamefont {Mariani}, \citenamefont {Minazzoli}, \citenamefont {Fienga}, \citenamefont {Laskar},\ and\ \citenamefont {Gastineau}}]{Mariani:2023rca}%
  \BibitemOpen
  \bibfield  {author} {\bibinfo {author} {\bibfnamefont {V.}~\bibnamefont {Mariani}}, \bibinfo {author} {\bibfnamefont {O.}~\bibnamefont {Minazzoli}}, \bibinfo {author} {\bibfnamefont {A.}~\bibnamefont {Fienga}}, \bibinfo {author} {\bibfnamefont {J.}~\bibnamefont {Laskar}},\ and\ \bibinfo {author} {\bibfnamefont {M.}~\bibnamefont {Gastineau}},\ }\bibfield  {title} {\bibinfo {title} {{Bayesian test of Brans\textendash{}Dicke theories with planetary ephemerides: Investigating the strong equivalence principle}},\ }\href {https://doi.org/10.1051/0004-6361/202348082} {\bibfield  {journal} {\bibinfo  {journal} {Astron. Astrophys.}\ }\textbf {\bibinfo {volume} {682}},\ \bibinfo {pages} {A175} (\bibinfo {year} {2024})},\ \Eprint {https://arxiv.org/abs/2310.00719} {arXiv:2310.00719 [astro-ph.EP]} \BibitemShut {NoStop}%
\bibitem [{\citenamefont {{Titov}}\ \emph {et~al.}(2018)\citenamefont {{Titov}}, \citenamefont {{Girdiuk}}, \citenamefont {{Lambert}}, \citenamefont {{Lovell}}, \citenamefont {{McCallum}}, \citenamefont {{Shabala}}, \citenamefont {{McCallum}}, \citenamefont {{Mayer}}, \citenamefont {{Schartner}}, \citenamefont {{de Witt}}, \citenamefont {{Shu}}, \citenamefont {{Melnikov}}, \citenamefont {{Ivanov}}, \citenamefont {{Mikhailov}}, \citenamefont {{Yi}}, \citenamefont {{Soja}}, \citenamefont {{Xia}},\ and\ \citenamefont {{Jiang}}}]{2018A&A...618A...8T}%
  \BibitemOpen
  \bibfield  {author} {\bibinfo {author} {\bibfnamefont {O.}~\bibnamefont {{Titov}}}, \bibinfo {author} {\bibfnamefont {A.}~\bibnamefont {{Girdiuk}}}, \bibinfo {author} {\bibfnamefont {S.~B.}\ \bibnamefont {{Lambert}}}, \bibinfo {author} {\bibfnamefont {J.}~\bibnamefont {{Lovell}}}, \bibinfo {author} {\bibfnamefont {J.}~\bibnamefont {{McCallum}}}, \bibinfo {author} {\bibfnamefont {S.}~\bibnamefont {{Shabala}}}, \bibinfo {author} {\bibfnamefont {L.}~\bibnamefont {{McCallum}}}, \bibinfo {author} {\bibfnamefont {D.}~\bibnamefont {{Mayer}}}, \bibinfo {author} {\bibfnamefont {M.}~\bibnamefont {{Schartner}}}, \bibinfo {author} {\bibfnamefont {A.}~\bibnamefont {{de Witt}}}, \bibinfo {author} {\bibfnamefont {F.}~\bibnamefont {{Shu}}}, \bibinfo {author} {\bibfnamefont {A.}~\bibnamefont {{Melnikov}}}, \bibinfo {author} {\bibfnamefont {D.}~\bibnamefont {{Ivanov}}}, \bibinfo {author} {\bibfnamefont {A.}~\bibnamefont {{Mikhailov}}}, \bibinfo {author} {\bibfnamefont {S.}~\bibnamefont {{Yi}}}, \bibinfo {author}
  {\bibfnamefont {B.}~\bibnamefont {{Soja}}}, \bibinfo {author} {\bibfnamefont {B.}~\bibnamefont {{Xia}}},\ and\ \bibinfo {author} {\bibfnamefont {T.}~\bibnamefont {{Jiang}}},\ }\bibfield  {title} {\bibinfo {title} {{Testing general relativity with geodetic VLBI. What a single, specially designed experiment can teach us}},\ }\href {https://doi.org/10.1051/0004-6361/201833459} {\bibfield  {journal} {\bibinfo  {journal} {\aap}\ }\textbf {\bibinfo {volume} {618}},\ \bibinfo {eid} {A8} (\bibinfo {year} {2018})},\ \Eprint {https://arxiv.org/abs/1806.11299} {arXiv:1806.11299 [astro-ph.IM]} \BibitemShut {NoStop}%
\bibitem [{\citenamefont {De~Marchi}\ and\ \citenamefont {Cascioli}(2020)}]{DeMarchi:2019lei}%
  \BibitemOpen
  \bibfield  {author} {\bibinfo {author} {\bibfnamefont {F.}~\bibnamefont {De~Marchi}}\ and\ \bibinfo {author} {\bibfnamefont {G.}~\bibnamefont {Cascioli}},\ }\bibfield  {title} {\bibinfo {title} {{Testing General Relativity in the Solar System: present and future perspectives}},\ }\href {https://doi.org/10.1088/1361-6382/ab6ae0} {\bibfield  {journal} {\bibinfo  {journal} {Class. Quant. Grav.}\ }\textbf {\bibinfo {volume} {37}},\ \bibinfo {pages} {095007} (\bibinfo {year} {2020})},\ \Eprint {https://arxiv.org/abs/1911.05561} {arXiv:1911.05561 [gr-qc]} \BibitemShut {NoStop}%
\bibitem [{\citenamefont {Bolton}\ \emph {et~al.}(2006)\citenamefont {Bolton}, \citenamefont {Rappaport},\ and\ \citenamefont {Burles}}]{Bolton:2006yz}%
  \BibitemOpen
  \bibfield  {author} {\bibinfo {author} {\bibfnamefont {A.~S.}\ \bibnamefont {Bolton}}, \bibinfo {author} {\bibfnamefont {S.}~\bibnamefont {Rappaport}},\ and\ \bibinfo {author} {\bibfnamefont {S.}~\bibnamefont {Burles}},\ }\bibfield  {title} {\bibinfo {title} {{Constraint on the Post-Newtonian Parameter gamma on Galactic Size Scales}},\ }\href {https://doi.org/10.1103/PhysRevD.74.061501} {\bibfield  {journal} {\bibinfo  {journal} {Phys. Rev. D}\ }\textbf {\bibinfo {volume} {74}},\ \bibinfo {pages} {061501} (\bibinfo {year} {2006})},\ \Eprint {https://arxiv.org/abs/astro-ph/0607657} {arXiv:astro-ph/0607657} \BibitemShut {NoStop}%
\bibitem [{\citenamefont {{Schwab}}\ \emph {et~al.}(2010)\citenamefont {{Schwab}}, \citenamefont {{Bolton}},\ and\ \citenamefont {{Rappaport}}}]{2010ApJ...708..750S}%
  \BibitemOpen
  \bibfield  {author} {\bibinfo {author} {\bibfnamefont {J.}~\bibnamefont {{Schwab}}}, \bibinfo {author} {\bibfnamefont {A.~S.}\ \bibnamefont {{Bolton}}},\ and\ \bibinfo {author} {\bibfnamefont {S.~A.}\ \bibnamefont {{Rappaport}}},\ }\bibfield  {title} {\bibinfo {title} {{Galaxy-Scale Strong-Lensing Tests of Gravity and Geometric Cosmology: Constraints and Systematic Limitations}},\ }\href {https://doi.org/10.1088/0004-637X/708/1/750} {\bibfield  {journal} {\bibinfo  {journal} {\apj}\ }\textbf {\bibinfo {volume} {708}},\ \bibinfo {pages} {750} (\bibinfo {year} {2010})},\ \Eprint {https://arxiv.org/abs/0907.4992} {arXiv:0907.4992 [astro-ph.CO]} \BibitemShut {NoStop}%
\bibitem [{\citenamefont {Guerrini}\ and\ \citenamefont {M\"ortsell}(2024)}]{Guerrini:2023pre}%
  \BibitemOpen
  \bibfield  {author} {\bibinfo {author} {\bibfnamefont {S.}~\bibnamefont {Guerrini}}\ and\ \bibinfo {author} {\bibfnamefont {E.}~\bibnamefont {M\"ortsell}},\ }\bibfield  {title} {\bibinfo {title} {{Probing a scale dependent gravitational slip with galaxy strong lensing systems}},\ }\href {https://doi.org/10.1103/PhysRevD.109.023533} {\bibfield  {journal} {\bibinfo  {journal} {Phys. Rev. D}\ }\textbf {\bibinfo {volume} {109}},\ \bibinfo {pages} {023533} (\bibinfo {year} {2024})},\ \Eprint {https://arxiv.org/abs/2309.11915} {arXiv:2309.11915 [astro-ph.CO]} \BibitemShut {NoStop}%
\bibitem [{\citenamefont {Liu}\ and\ \citenamefont {Liao}(2024)}]{Liu:2023ulr}%
  \BibitemOpen
  \bibfield  {author} {\bibinfo {author} {\bibfnamefont {T.}~\bibnamefont {Liu}}\ and\ \bibinfo {author} {\bibfnamefont {K.}~\bibnamefont {Liao}},\ }\bibfield  {title} {\bibinfo {title} {{Determining cosmological-model-independent H0 and post-Newtonian parameter with time-delay lenses and supernovae}},\ }\href {https://doi.org/10.1093/mnras/stae119} {\bibfield  {journal} {\bibinfo  {journal} {Mon. Not. Roy. Astron. Soc.}\ }\textbf {\bibinfo {volume} {528}},\ \bibinfo {pages} {1354} (\bibinfo {year} {2024})},\ \Eprint {https://arxiv.org/abs/2309.13608} {arXiv:2309.13608 [astro-ph.CO]} \BibitemShut {NoStop}%
\bibitem [{\citenamefont {Liu}\ \emph {et~al.}(2024)\citenamefont {Liu}, \citenamefont {Biesiada}, \citenamefont {Tian},\ and\ \citenamefont {Liao}}]{Liu:2024bre}%
  \BibitemOpen
  \bibfield  {author} {\bibinfo {author} {\bibfnamefont {T.}~\bibnamefont {Liu}}, \bibinfo {author} {\bibfnamefont {M.}~\bibnamefont {Biesiada}}, \bibinfo {author} {\bibfnamefont {S.}~\bibnamefont {Tian}},\ and\ \bibinfo {author} {\bibfnamefont {K.}~\bibnamefont {Liao}},\ }\bibfield  {title} {\bibinfo {title} {{Robust test of general relativity at the galactic scales by combining strong lensing systems and gravitational wave standard sirens}},\ }\href {https://doi.org/10.1103/PhysRevD.109.084074} {\bibfield  {journal} {\bibinfo  {journal} {Phys. Rev. D}\ }\textbf {\bibinfo {volume} {109}},\ \bibinfo {pages} {084074} (\bibinfo {year} {2024})},\ \Eprint {https://arxiv.org/abs/2404.05907} {arXiv:2404.05907 [gr-qc]} \BibitemShut {NoStop}%
\bibitem [{\citenamefont {Wei}\ \emph {et~al.}(2022)\citenamefont {Wei}, \citenamefont {Chen}, \citenamefont {Cao},\ and\ \citenamefont {Wu}}]{Wei:2022rcb}%
  \BibitemOpen
  \bibfield  {author} {\bibinfo {author} {\bibfnamefont {J.-J.}\ \bibnamefont {Wei}}, \bibinfo {author} {\bibfnamefont {Y.}~\bibnamefont {Chen}}, \bibinfo {author} {\bibfnamefont {S.}~\bibnamefont {Cao}},\ and\ \bibinfo {author} {\bibfnamefont {X.-F.}\ \bibnamefont {Wu}},\ }\bibfield  {title} {\bibinfo {title} {{Direct Estimate of the Post-Newtonian Parameter and Cosmic Curvature from Galaxy-scale Strong Gravitational Lensing}},\ }\href {https://doi.org/10.3847/2041-8213/ac551e} {\bibfield  {journal} {\bibinfo  {journal} {Astrophys. J. Lett.}\ }\textbf {\bibinfo {volume} {927}},\ \bibinfo {pages} {L1} (\bibinfo {year} {2022})},\ \Eprint {https://arxiv.org/abs/2202.07860} {arXiv:2202.07860 [astro-ph.CO]} \BibitemShut {NoStop}%
\bibitem [{\citenamefont {{Koopmans}}(2006)}]{2006EAS....20..161K}%
  \BibitemOpen
  \bibfield  {author} {\bibinfo {author} {\bibfnamefont {L.~V.~E.}\ \bibnamefont {{Koopmans}}},\ }\bibfield  {title} {\bibinfo {title} {{Gravitational Lensing \& Stellar Dynamics}},\ }in\ \href {https://doi.org/10.1051/eas:2006064} {\emph {\bibinfo {booktitle} {EAS Publications Series}}},\ \bibinfo {series} {EAS Publications Series}, Vol.~\bibinfo {volume} {20},\ \bibinfo {editor} {edited by\ \bibinfo {editor} {\bibfnamefont {G.~A.}\ \bibnamefont {{Mamon}}}, \bibinfo {editor} {\bibfnamefont {F.}~\bibnamefont {{Combes}}}, \bibinfo {editor} {\bibfnamefont {C.}~\bibnamefont {{Deffayet}}},\ and\ \bibinfo {editor} {\bibfnamefont {B.}~\bibnamefont {{Fort}}}}\ (\bibinfo {year} {2006})\ pp.\ \bibinfo {pages} {161--166},\ \Eprint {https://arxiv.org/abs/astro-ph/0511121} {arXiv:astro-ph/0511121 [astro-ph]} \BibitemShut {NoStop}%
\bibitem [{\citenamefont {Chen}\ \emph {et~al.}(2019)\citenamefont {Chen}, \citenamefont {Li}, \citenamefont {Shu},\ and\ \citenamefont {Cao}}]{Chen:2018jcf}%
  \BibitemOpen
  \bibfield  {author} {\bibinfo {author} {\bibfnamefont {Y.}~\bibnamefont {Chen}}, \bibinfo {author} {\bibfnamefont {R.}~\bibnamefont {Li}}, \bibinfo {author} {\bibfnamefont {Y.}~\bibnamefont {Shu}},\ and\ \bibinfo {author} {\bibfnamefont {X.}~\bibnamefont {Cao}},\ }\bibfield  {title} {\bibinfo {title} {{Assessing the effect of lens mass model in cosmological application with updated galaxy-scale strong gravitational lensing sample}},\ }\href {https://doi.org/10.1093/mnras/stz1902} {\bibfield  {journal} {\bibinfo  {journal} {Mon. Not. Roy. Astron. Soc.}\ }\textbf {\bibinfo {volume} {488}},\ \bibinfo {pages} {3745} (\bibinfo {year} {2019})},\ \Eprint {https://arxiv.org/abs/1809.09845} {arXiv:1809.09845 [astro-ph.CO]} \BibitemShut {NoStop}%
\bibitem [{\citenamefont {Robitaille}\ \emph {et~al.}(2013)\citenamefont {Robitaille} \emph {et~al.}}]{Astropy:2013muo}%
  \BibitemOpen
  \bibfield  {author} {\bibinfo {author} {\bibfnamefont {T.~P.}\ \bibnamefont {Robitaille}} \emph {et~al.} (\bibinfo {collaboration} {Astropy}),\ }\bibfield  {title} {\bibinfo {title} {{Astropy: A Community Python Package for Astronomy}},\ }\href {https://doi.org/10.1051/0004-6361/201322068} {\bibfield  {journal} {\bibinfo  {journal} {Astron. Astrophys.}\ }\textbf {\bibinfo {volume} {558}},\ \bibinfo {pages} {A33} (\bibinfo {year} {2013})},\ \Eprint {https://arxiv.org/abs/1307.6212} {arXiv:1307.6212 [astro-ph.IM]} \BibitemShut {NoStop}%
\bibitem [{\citenamefont {van~der Walt}\ \emph {et~al.}(2011)\citenamefont {van~der Walt}, \citenamefont {Colbert},\ and\ \citenamefont {Varoquaux}}]{vanderWalt:2011bqk}%
  \BibitemOpen
  \bibfield  {author} {\bibinfo {author} {\bibfnamefont {S.}~\bibnamefont {van~der Walt}}, \bibinfo {author} {\bibfnamefont {S.~C.}\ \bibnamefont {Colbert}},\ and\ \bibinfo {author} {\bibfnamefont {G.}~\bibnamefont {Varoquaux}},\ }\bibfield  {title} {\bibinfo {title} {{The NumPy Array: A Structure for Efficient Numerical Computation}},\ }\href {https://doi.org/10.1109/MCSE.2011.37} {\bibfield  {journal} {\bibinfo  {journal} {Comput. Sci. Eng.}\ }\textbf {\bibinfo {volume} {13}},\ \bibinfo {pages} {22} (\bibinfo {year} {2011})},\ \Eprint {https://arxiv.org/abs/1102.1523} {arXiv:1102.1523 [cs.MS]} \BibitemShut {NoStop}%
\bibitem [{\citenamefont {Hunter}(2007)}]{Hunter:2007ouj}%
  \BibitemOpen
  \bibfield  {author} {\bibinfo {author} {\bibfnamefont {J.~D.}\ \bibnamefont {Hunter}},\ }\bibfield  {title} {\bibinfo {title} {{Matplotlib: A 2D Graphics Environment}},\ }\href {https://doi.org/10.1109/MCSE.2007.55} {\bibfield  {journal} {\bibinfo  {journal} {Comput. Sci. Eng.}\ }\textbf {\bibinfo {volume} {9}},\ \bibinfo {pages} {90} (\bibinfo {year} {2007})}\BibitemShut {NoStop}%
\bibitem [{\citenamefont {Virtanen}\ \emph {et~al.}(2020)\citenamefont {Virtanen} \emph {et~al.}}]{Virtanen:2019joe}%
  \BibitemOpen
  \bibfield  {author} {\bibinfo {author} {\bibfnamefont {P.}~\bibnamefont {Virtanen}} \emph {et~al.},\ }\bibfield  {title} {\bibinfo {title} {{SciPy 1.0--Fundamental Algorithms for Scientific Computing in Python}},\ }\href {https://doi.org/10.1038/s41592-019-0686-2} {\bibfield  {journal} {\bibinfo  {journal} {Nature Meth.}\ }\textbf {\bibinfo {volume} {17}},\ \bibinfo {pages} {261} (\bibinfo {year} {2020})},\ \Eprint {https://arxiv.org/abs/1907.10121} {arXiv:1907.10121 [cs.MS]} \BibitemShut {NoStop}%
\bibitem [{\citenamefont {McKinney}(2010)}]{mckinney-proc-scipy-2010}%
  \BibitemOpen
  \bibfield  {author} {\bibinfo {author} {\bibfnamefont {W.}~\bibnamefont {McKinney}},\ }\bibfield  {title} {\bibinfo {title} {Data structures for statistical computing in python},\ }in\ \href@noop {} {\emph {\bibinfo {booktitle} {Proceedings of the 9th Python in Science Conference}}},\ \bibinfo {editor} {edited by\ \bibinfo {editor} {\bibfnamefont {S.}~\bibnamefont {van~der Walt}}\ and\ \bibinfo {editor} {\bibfnamefont {J.}~\bibnamefont {Millman}}}\ (\bibinfo {year} {2010})\ pp.\ \bibinfo {pages} {51 -- 56}\BibitemShut {NoStop}%
\bibitem [{\citenamefont {Waskom}(2021)}]{Waskom2021}%
  \BibitemOpen
  \bibfield  {author} {\bibinfo {author} {\bibfnamefont {M.~L.}\ \bibnamefont {Waskom}},\ }\bibfield  {title} {\bibinfo {title} {seaborn: statistical data visualization},\ }\href {https://doi.org/10.21105/joss.03021} {\bibfield  {journal} {\bibinfo  {journal} {Journal of Open Source Software}\ }\textbf {\bibinfo {volume} {6}},\ \bibinfo {pages} {3021} (\bibinfo {year} {2021})}\BibitemShut {NoStop}%
\bibitem [{\citenamefont {{Foreman-Mackey}}\ \emph {et~al.}(2013)\citenamefont {{Foreman-Mackey}}, \citenamefont {{Hogg}}, \citenamefont {{Lang}},\ and\ \citenamefont {{Goodman}}}]{2013PASP..125..306F}%
  \BibitemOpen
  \bibfield  {author} {\bibinfo {author} {\bibfnamefont {D.}~\bibnamefont {{Foreman-Mackey}}}, \bibinfo {author} {\bibfnamefont {D.~W.}\ \bibnamefont {{Hogg}}}, \bibinfo {author} {\bibfnamefont {D.}~\bibnamefont {{Lang}}},\ and\ \bibinfo {author} {\bibfnamefont {J.}~\bibnamefont {{Goodman}}},\ }\bibfield  {title} {\bibinfo {title} {{emcee: The MCMC Hammer}},\ }\href {https://doi.org/10.1086/670067} {\bibfield  {journal} {\bibinfo  {journal} {\pasp}\ }\textbf {\bibinfo {volume} {125}},\ \bibinfo {pages} {306} (\bibinfo {year} {2013})},\ \Eprint {https://arxiv.org/abs/1202.3665} {arXiv:1202.3665 [astro-ph.IM]} \BibitemShut {NoStop}%
\bibitem [{\citenamefont {Will}(1971{\natexlab{b}})}]{Will:1971zzb}%
  \BibitemOpen
  \bibfield  {author} {\bibinfo {author} {\bibfnamefont {C.~M.}\ \bibnamefont {Will}},\ }\bibfield  {title} {\bibinfo {title} {{Theoretical Frameworks for Testing Relativistic Gravity. 2. Parametrized Post-Newtonian Hydrodynamics, and the Nordtvedt Effect}},\ }\href {https://doi.org/10.1086/150804} {\bibfield  {journal} {\bibinfo  {journal} {Astrophys. J.}\ }\textbf {\bibinfo {volume} {163}},\ \bibinfo {pages} {611} (\bibinfo {year} {1971}{\natexlab{b}})}\BibitemShut {NoStop}%
\end{thebibliography}%

\end{document}